# Einstein's Fluctuation Formula. A Historical Overview

**Sándor Varró**

Research Institute for Solid State Physics and Optics
Hungarian Academi of Sciences
Letters: H-1525 Budapest, POBox 49, Hungary
E-mail: varro@sunserv.kfki.hu

**Abstract.** A historical overview is given on the basic results which appeared by the year 1926 concerning Einstein's fluctuation formula of black-body radiation, in the context of light-quanta and wave-particle duality. On the basis of the original publications – from Planck's derivation of the black-body spectrum and Einstein's introduction of the photons up to the results of Born, Heisenberg and Jordan on the quantization of a continuum – a comparative study is presented on the first lines of thoughts that led to the concept of quanta. The nature of the particle-like fluctuations and the wave-like fluctuations are analysed by using several approaches. With the help of classical probability theory, it is shown that the infinite divisibility of the Bose distribution leads to the new concept of classical "poissonian photo-multiplets" or to the  "binary photo-multiplets" of fermionic character. As an application, Einstein's fluctuation formula is derived as a sum of fermion type fluctuations of the binary photo-multiplets.

## 1. Introduction

Nearly one hundred years ago, in 1909, Albert Einstein published his famous paper [1] entitled "*Zum gegenwärtigen Stand des Strahlungsproblems*" ("*To the recent state of art of the radiation problem*"), where he gave the first mathematically correct formula expressing the "wave-particle duality" in the case of black-body radiation. This formula is called Einstein's fluctuation formula. It gives the variance (mean square deviation) of the energy of black-body radiation in a narrow spectral range in a sub-volume of a cavity surrounded by perfectly reflecting walls (a "*Hohlraum*" in the German terminology). The formula contains two terms, the "particle-term" (the Wien term) and the "wave-term" (the Rayleigh-Jeans term). The two terms simply add, as if they were stemming from independent processes.

Let us now imagine the photons as point-like particles, then it is clear that in thermal equilibrium they fill up the cavity homogeneously on average, but their actual number in any sub-volume will vary by chance. Thus the actual number of the photons is a random variable (following the Poisson distribution of "rare events") whose variance governs the energy fluctuation in the sub-volume. Since in this case the variance of the photon number equals to its expectation value, the energy fluctuation is given by the product of *one* photon energy and the average energy contained in the sub-volume. The "particle-term" of the fluctuation formula has the *same form*, but its numerical value is different.

On the other hand, we can consider the heat radiation as a superposition of electromagnetic waves (the eigenmodes of the cavity), an oscillating continuum with random phases. Then, though the average of the amplitude of a spectral component is zero, the square of the amplitude (which is proportional to the electromagnetic energy density) will have in average a certain finite, uniform value determined by the temperature alone. The wave-like fluctuations in a sub-volume of the cavity can be imagined as a result of an irregular "breeding" of the beat waves formed from the interfering Fourier components. Classical analysis shows that this "wave-term" (in a narrow spectral range) equals to the ratio of the average energy squared to the number of modes in the sub-volume. The second term in Einstein's fluctuation formula has the *same form*, but again its numerical value is different, like in the case of the particle fluctuations.

In Einstein's fluctuation formula the two terms are equal if the average occupation number of the modes is unity. In this case the photon energy $h\nu=kT\log 2$ just equals to the minimum amount of energy necessary to transmit one bit of information according to Shannon. For small densities of the black-body radiation ($h\nu/kT \gg 1$) the particle-term (Wien-term) dominates, and for large radiation densities ($h\nu/kT \ll 1$) the wave-term (Rayleigh-Jeans-term) dominates ($k$ denotes the Boltzmann constant and $T$ is the absolute temperature). From the conceptual point of view, the importance of Einstein's fluctuation formula appears in the contex of the introduction of photons. Einstein's original analysis [3] was based on the study of the entropy of black-body radiation in the Wien limit: "Monochromatic radiation of small density (in the range of the validity of Wien's radiation formula) behaves so from the point of view of heat theory as if it consisted of independent energy quanta …". The derivation that led him to this conclusion does not work if one uses the exact Planck formula instead of the *approximate* Wien formula, so this way the photon concept could not be justified generally. However, his derivation of the fluctuation formula four years later was based on Planck's *exact* expression, and it contains the particle-like fluctuation which may be identified with the fluctuation of the number of photons. In this way the photon concept received a firm support.

We may safely state that all the results of Einstein concerning the photons (light-quanta, as he called them) were based exclusively on the study of black-body radiation. That is why in Section 2 we have to start with a summary on the basics of black-body radiation, includig

Planck's original derivation of the correct spectral distribution, and the discovery of the elementary quantum of action. In Section 3 Einstein's argumentation is presented which led him to introduce the concept of light quanta. The derivation of the fluctuation formula is given in two different ways. Then the physical content of the fluctuation formula is analysed on the basis of classical probability theory. In Section 4 an early derivation of the Bose distribution is discussed in the framework of a general combinatorial analysis, and then a part of Planck's studies on the fluctuations are summarized briefly. In Section 5 the results of Ehrenfest and Smekal are reviewed. The superposition of classical random waves with discrete energy distributions, and the effect of a ponderable material particle on the fluctuations in the *Hohlraum* are discussed. The second part of Section 5 is devoted to the work of Born, Heisenberg and Jordan in which the first formulation of the field quantization appeared, where the field amplitudes were represented by matrices (operators). In Section 6 our recent new results on the infinite divisibility of the Bose distribution are presented and Einstein's fluctuation formula is derived from pure particle-like and from fermion type fluctuations. Section 7 closes the paper with a brief summary. Some of the derivations and interpretations to be presented here will surely be well known to the reader, because just these results have been widely accepted in the meantime and become basic parts of textbooks on the quantum theory of light or on quantum electronics. In our opinion, it is always very useful to get aquaintance with the development of original ideas and methods – even if some had led dead ends or detours – which finally led to the creation of the clean structure of today's canonical science. Einstein's fluctuation formula can be derived in two lines if one uses the Bose distribution or the creation and annihilation operators of the photons. On the other hand, the physical content of it, and the nature of the fluctuation of a *real* black-body radiation do not come out automatically from the formalism. In writing up the present paper – following the original paths of the great masters – our main motivation was to give a deeper insight to the physical significance of the fluctuation formula from the point of view of the wave-particle duality which still puzzles many physicists including us. Finally, some technical remarks: the calculations are presented in the main text, on purpose, because just these details show clearly the *differentia specifica* of the particular approaches. Though we have tried to make the paper a coherent unit, usually the original notations are also used separately, and the Sections are essentially self-contained.

**2. Planck's Law of Black-Body Radiation**

**2.1. *The Stefan-Boltzmann law and Wien's displacement law***
The heat radiation of black-bodies has certain universal characters whose study has led Max Planck in 1900 to find – besides the correct spectral distribution of this radiation – the new universal constant $h=6.626 \times 10^{-27} erg.sec$, the elementary quantum of action [2], which plays a fundamental role not only in quantum physics but also in our everyday life. Because five years later Einstein introduced his hypothesis of light quanta [3] on the basis of the thermodynamical analysis of black-body radiation, and he has returned from time to time to this subject in his investigations on light quanta, we cannot get around Planck's path-breaking work in the present paper, at least up to a short summary [4].
Everyday experience shows and classical physics tells us that any material body of a finite temperature emits electromagnetic radiation, and it also absorbs such heat radiation from its surroundings. In general, this radiation consist of infinitely many components of different frequencies and of two independent polarizations. According to Kirchhoff, when a body is in thermal equilibrium with its surroundings, and moreover its material is homogeneous and isotropic, then all over in the inner part of the body and on its surface, too, the ratios of the emission and the absorption capabilities belonging to these spectral components are independent of the material constitution of the body, and they are equal to the emission capability of an absolutely black body. This latter one depends only on the absolute temperature $T$ and on the frequency $\nu$. The spectral density $u_\nu$ – which is the fraction of the radiant energy in the frequency interval ($\nu, \nu+d\nu$) in a unit volume – may depend on the material constitution. However, on the basis of general considerations [4] Clausius has shown that by passing through the common interface of two bodies $K$ and $K'$ being in thermal equilibrium, there is an invariant $c^3 u_\nu = c'^3 u'_\nu$, where $c$ and $c'$ are the corresponding velocities of propagation of the radiation. Hence, if a black body is in thermal equilibrium with a thermal radiation in vacuum closed in a cavity with perfectly reflecting walls, then $u_\nu$ must be an universal function $u_\nu = u(\nu, T)$, because now $c$ is a universal constant, namely the velocity of light *in vacuo*. Moreover, it can be proved [4] that in a medium being transparent for a given color, such a component of the radiation can be in thermal equilibrium with its surroundings at an arbitrary intensity. *Consequently, in a region of vacuum surrounded by perfectly reflecting boundaries (in a "Hohlraum", i.e. in an empty cavity) the radiation can be in thermal equilibrium in any state, but these equilibria are not stable in general. When we place into the cavity a small piece of a ponderable matter (Planck's "Kohlenstäubchen", a small stick of carbon capable of absorbing all the spectral components of the radiation), then, in the course of proceeding to the a newer equilibrium, the spectral distribution will be rearranged to the spectrum of the black-body radiation.* During this rearrangement the total energy of the radiant heat will not change, the small carbon stick merely plays the role of initiating the process. This is similar, for example to the condensation of an overdense vapor which is initialized by a small liquid droplet, and the system gets to a stable state of maximum entropy with practically no net energy change. According to the above considerations, such a heat radiation closed in the cavity is a *black-body radiation whose spectral density $u_\nu = u(\nu, T)$ is a universal function of the frequency and of the absolute temperature. In the general case there belongs a spectral entropy density $s_\nu$, and consequently an absolute temperature $(\partial s_\nu / \partial u_\nu)^{-1} = T_\nu$ to each component of the cavity radiation [4]. The black-body radiation is just characterized by that each spectral components are at that same temperature.* The last two results of classical physics being in quantitative agreement with the experiments are the Stefan-Boltzmann law (1879, 1884) and Wien's displacement law (1893), each of which can be deduced by using thermodynamic considerations and results of classical electrodynamics concerning radiation pressure and the Doppler effect [4]. It is interesting to mention that the displacement law had already been published eight years earlier in 1885 by R. von Kövesligethy [5] who used the mechanical concept of aether in the derivation of his spectral formula.

The temperature dependence of the total energy density $u(T) = \int u(\nu, T) d\nu \sim T^4$ was first found by Joseph Stefan in 1879 as an empirical formula on the basis of quite inaccurate measurements. Then in 1884 Ludwig Boltzmann confirmed this formula by an exact derivation based on phenomenological thermodynamics. Now we are going to summarize the basic steps of this simple derivation. If the radiation, being enclosed in a cavity of volume $V$ and of perfectly reflecting walls, undergoes an isoterm expansion by pushing slowly outwards a piston, then, according to the first law of thermodynamics, the surroundings gives the amount of heat $Q = d(uV) + (u/3)dV$ to the cavity. Here we have taken into account that, according to the Maxwell equations, the radiation pressure is one third of the energy density in this case. The entropy change during this process becomes $dS(V,T) = Q/T = (V/T)(du/dT)dT + (4u/3T)dV = (\partial S/\partial T)_V dT + (\partial S/\partial V)_T dV$. From the equality of the second order partial differentials, $(\partial^2 S/\partial V \partial T) = (\partial^2 S/\partial T \partial V)$, we obtain the differential equation $du/dT = 4u/T$, whose solution reads $u = \sigma T^4$, where $\sigma$ is called the Stefan-Boltzmann constant.

Wilhelm Wien used the following line of thought in deriving his displacement law [6]. If we assume that the heat radiation fills a slowly shrinking sphere of instantaneous radius $r$, then the wavelengths of every spectral components undergo a Doppler shift and they follow linearly the change in $r$, that is $c/\nu = \lambda = const \times r$. The energy density $E(\lambda,T)d\lambda = u(\nu,T)d\nu$ falling in the spectral range $(\lambda, \lambda+d\lambda)$ will then be inversely proportional with the fourth power of the radius $E(\lambda,T)d\lambda = const/r^4$. On the other hand, according to the Stefan-Boltmann law, we have for the integrated energy density $\int E(\lambda,T)d\lambda = const/r^4 = \sigma T^4$, which means that the absolut temperature is inversely proportional with the slowly varying radius, i.e. $T = const/r$. We note that expression "slowly varying" here means that we neglect terms of order $\dot{r}^2/c^2$ in calculating the Doppler shift. On the basis of these results if we compare two arbitrary states of the system, we have $[E(\lambda_1,T_1)d\lambda_1]/[E(\lambda_2,T_2)d\lambda_2] = r_2^4/r_1^4 \equiv q^4$, that is $E(\lambda,T)d\lambda = q^4 E(q\lambda, q^{-1}T)d(q\lambda)$. The solution of the latter functional equation can be expressed as $E(\lambda,T) = \lambda^{-5}\Phi(\lambda T)$. The dependence on the frequency reads $u(\nu,T) = \nu^3 F(\nu/T)$, where $\Phi$ and $F$ are universal functions. From the condition $[\partial E/\partial \lambda]_{max} = 0$ we obtain that $\lambda_{max} T = [5\Phi/\Phi']_{max} = const$ is a universal constant whose experimental value is $0.2899 cm \times grad$. According to Wien's displacement law, by increasing the temperature of the black-body radiation the position of the maximum energy density is shifted towards the shorter wavelengths (larger frequencies).

**2.2. *The introduction of Planck's elementary quantum of action***
In spite of very much effort using classical physics, no correct formula had been found by the year 1900 for the frequency dependence of the spectral energy density $u_\nu$ which would have had given back the dependence in the whole frequency range measured with a very high precision by O. Lummer and E. Pringsheim [7] and H. Rubens and F. Kurlbaum [8]. In 1896 Wien published a formula [9] for $u_\nu$ which worked quite well over the large frequency wing of the spectrum, but for long wavelengths it failed to give back the experimental results by Rubens and Kurlbaum. This was just the formula which was remedied by Planck with the help of a "fortunate interpolation" and led to the correct spectral equation [10]. The spectral formula derived by Lord Rayleigh in 1900, and recovered later on a different basis by James Jeans [11] fits well to the experimental data for long wavelengths, but gives infinite total energy after integrating with respect to the frequencies from zero to infinity. This artifact is called the "ultraviolet catastrophe".
Because the spectrum of a black-body radiation does not depend on the material constitution of the body with which it is in thermal equilibrium, we are allowed to model the material system at will. Planck has chosen an assembly of linear oscillators whose eigenfrequencies covered the whole spectrum, such that they could get into resonance with all the components of the radiation. By applying classical electrodynamics he proved first [12] that the spectral density of the radiation being in equilibrium with the resonators is given by $u_\nu(\nu,T) = (8\pi\nu^2/c^3)U_1(\nu,T)$, where $U_1(\nu,T)$ denotes the *average energy of one oscillator*. It is interesting to note that $(8\pi\nu^2/c^3) = Z_\nu$ is just the spectral mode density of the radiation when the linear extensions of the cavity are much larger than the wavelengths considered. Planck got the correct determination of the quantity $U_1$ through the study of the entropy of the oscillator manyfold [2] in the following way. It is clear that the average energy of $N$ oscillators is $U_N = NU_1$, and a similar relation holds for the corresponding entropy, $S_N = NS_1$. According to Boltzmann's principle the entropy corresponding to a "macrostate" of the ensemble of the oscillators can be expressed as $S_N = k\log W_N$, where $W_N$ denotes the number of all those "microstates" or "complexions" which belong to the same total energy $U_N$. Planck's revolutionary new idea was that he did not consider the total energy as an infinitely divisible continuous quantity, but he assumed that it consists of energy quanta of finite size $\varepsilon$ and of finite number P, that is $NU_1=P\varepsilon$. As he wrote on page 556 of Ref. [2][1] : "*It comes about to find the probability W of that the N resonators altogether possess the oscillation energy $U_N$. To this end it is necessary to think of $U_N$ as not being a continuous, unlimitedly divisible quantity, but rather a discrete quantity built up of a finite number of identical parts. When we call such a part an energy element $\varepsilon$, then we have to set : $U_N = P\varepsilon$, where P means an integer, generally a large number, and the value of $\varepsilon$ is still to be determined.*" The energy elements are exactly equal and cannot be distinguished from each other, so the number of ways they can be distributed among the N oscillators is given by the number of combinations with repetitions $W_{N,P}=N(N+1)...(N+P-1)/P!$, or $W_{N,P}=(N+P-1)!/(N-1)!P!$. By using Stirling's formula, $N! \approx (N/e)^N$, we obtain the expression for the entropy of one oscillator

$$S_1 = k[(1+U_1/\varepsilon)\log(1+U_1/\varepsilon) - (U_1/\varepsilon)\log(U_1/\varepsilon)], \qquad (1)$$

where we have used $P/N=U_1/\varepsilon$. On the basis of Wien's displacement law it can be proved [4] that $S_1$ must be of the form $S_1=f(U_1/\nu)$, where $f$ is an universal function, so, according to Eq. (1) $\varepsilon$ has to be proportional to the frequency, $\varepsilon=h\nu$. The constant of proportionality is just Planck's elementary quantum of action $h$ [2]. By using the fundamental relation $dS_1/dU_1=1/T$ of thermodynamics we can express the average energy $U_1$ from Eq. (1) as a function of the frequency and of the absolute temperature, and then through the equation $u_\nu(\nu,T)=(8\pi\nu^2/c^3)U_1(\nu,T)$ the spectral formula reads

$$U_1 = \frac{h\nu}{e^{h\nu/kT}-1} = \bar{n}h\nu, \quad \bar{n} = \frac{1}{e^{h\nu/kT}-1}, \qquad (2)$$

$$u_\nu = \frac{8\pi\nu^2}{c^3}\frac{h\nu}{e^{h\nu/kT}-1} = Z_\nu \bar{n}h\nu, \quad Z_\nu = \frac{8\pi\nu^2}{c^3}, \qquad (3)$$

where $\bar{n}$ denotes the mean number of quanta of an oscillator. By integrating the spectral density $u_\nu$ with respect to the frequency we obtain the Stefan-Boltzmann law $u=\sigma T^4$, where $\sigma=8\pi^5k^4/15c^3h^3$. When we compute the wavelength $\lambda_m$ at which the density $E(\lambda,T)=(c/\lambda^2)u(\nu=c/\lambda,T)$ takes its maximum we get the transcendental equation $e^{-b}+b/5-1=0$ with the numerical solution $b \equiv ch/k\lambda_m T=4.965...$ . From the experimental values of the parameter $\sigma$ of the Stefan-Boltzmann law and of the constant $\lambda_m T=0.2899$cm×grad of the Wien's displacement law one can calculate the Boltzmann constant $k=1.381\times 10^{-16}$erg/grad and the Planck constant $h=6.626\times 10^{-27}$erg.sec. The results embodied in Eqs. (1), (2) and (3) secure that the temperatures $(\partial s_\nu/\partial u_\nu)^{-1}=T_\nu$ are the same for all spectral components, hence $T_\nu$ does not depend on the frequency. This means that here we are really dealing with the black-body radiation in a stable equilibrium with maximum entropy.

In Planck's original line of thought *the energies of the resonators were quantized*, no word had been said about the quantization of the heat radiation itself, which was the essence of Einstein's hypotesis on light quanta five years later. This is partly why the nowadays widespread view has been accepted that, though it is confirmed 100% by the experience, Planck's derivation is conceptually incorrect. On the other hand, it is remarkable, that when we replace the number of oscillators by the number of modes $M=V(8\pi\nu^2/c^3)d\nu$ (the number of degrees of freedom in the frequency range ($\nu, \nu+d\nu$)) in the *Hohlraum* and, moreover, we reinterpret the average energy of one oscillator $U_1$ as the average energy of a particular mode, then we can repeat the derivation without any changes – as was done by Debye in 1910 – and arrive at the correct results expressed by Eqs. (1), (2) and (3). This way we receive an elegant and conceptionally correct procedure, which is very simple at the same time.

---

[1] The original German text on page 556 of Ref. [2] : "Es kommt nun darauf an, die Wahrscheinlichkeit *W* dafür zu finden, dass die *N* Resonatoren insgesamt die Schwingungsenergie $U_N$ besitzen. Hierzu ist es notwendig, $U_N$ nicht als eine stetige, unbeschränkt teilbare, sondern als eine discrete, aus einen ganzen Zahl von endlichen gleichen Teilen zusammengesetzte Grösse aufzufassen. Nennen wir einen solchen Teil ein Energieelement $\varepsilon$, so ist mithin zu setzen : $U_N = P\varepsilon$, wobei *P* eine ganze, im allgemein grosse Zahl bedeutet, während wir den Wert von $\varepsilon$ noch dahingestellt sein lassen."

Let us conclude the present section by deriving the Planck-Bose distribution with the help of Planck's original method. We are interested in the probability of the event when one particular mode is excited exactly to the *n*-th energy level. The other *M-1* modes have then *P-n* quanta which can be arranged in $W_{N-1,P-n}=(N-2+P-n)!/(N-2)!(P-n)!$ number of possible combinations. It is then natural to associate the probability $p_n = W_{N-1,P-n} / W_{N,P}$ to the *n*-th excitation of a mode. After a straightforward calculation we obtain

$$p_n = \frac{1}{1+\bar{n}}\left(\frac{\bar{n}}{1+\bar{n}}\right)^n, \quad \sum_{n=0}^{\infty} p_n = 1, \quad \bar{n} = \sum_{n=0}^{\infty} n p_n, \quad S_1 = -k\sum_{n=0}^{\infty} p_n \log p_n, \tag{4}$$

where $\bar{n}$ denotes the same mean occupation number which have already been introduced in Eq. (2). We call the discrete distribution { $p_n$ ; n=0, 1, 2, ... } given in Eq. (4) Planck-Bose distribution. It is the probability distribution of the energy quanta (excitations) belonging to one particular mode being in thermal equilibrium. One can check by direct computation that the Boltzmann entropy defined in the last equation of Eq. (4) coincides with the thermodynamical expression Eq. (1) due to Planck.

## 3. Einstein's Hyphotesis of Light Quanta and the Fluctuation Formula

### 3.1. *Einstein's hypothesis of light quanta*
Surely Einstein was the first among those who really took Planck's quantum hypothesis serious, and, at the same time, he used it to find the laws of the new physics at the beginning of the last century. In his first paper [3] appeared in 1905 on this subject he started the analysis of the black-body radiation with the following thoughts. According to Planck [2] the average energy of one resonator $U_1$ can be expressed by the spectral energy density of the radiation, $U_1(\nu,T)=u_\nu(\nu,T)(8\pi\nu^2/c^3)^{-1}$, as was already mentioned above. If now we applied the energy equipartition theorem of classical statistics, we would obtain $U_1=2\times(1/2)kT$, because the amount $(1/2)kT$ goes to both the kinetic and the potential energies of a linear oscillator. This way we would end up with the Rayleigh-Jeans law,
$u^{R-J}=(8\pi\nu^2/c^3)kT$, which is a good approximation if $h\nu/kT\ll 1$. For large frequencies, or for small radiation densities when $h\nu/kT\gg 1$, the Planck distribution goes over to the Wien limit $\rho$ (Einstein used the notation $\rho$ for the spectral density), and we can write

$$\rho = u^{Wien} = \alpha\nu^3 e^{-\beta\nu/T}, \text{ that is } 1/T = -(1/\beta\nu)\log(\rho/\alpha\nu^3), \tag{5}$$

where $\alpha=8\pi h/c^3$ and $\beta=h/k$. By taking into account the general relation of thermodynamics $\partial s/\partial \rho=1/T$ and integrating the second equation of Eq. (5) we obtain the expression for the spectral entropy density $s=-(\rho/\beta\nu)[log(\rho/\alpha\nu^3)-1]$. Let us now consider the heat radiation of total energy *E* in the spectral range ($\nu$, $\nu+d\nu$) distributed homogeneously over a volume *V*. Then $E=V\rho d\nu$ and the total entropy $S=Vsd\nu$ equals

$$S = Vsd\nu = -(E/\beta\nu)[\log(E/V\alpha\nu^3 d\nu) - 1]. \tag{6}$$

When this same radiation of energy *E* is distributed in a larger volume $V_0$, then the entropy difference of these two states is expressed from Eq. (6) by the following formula

$$S - S_0 = (E/\beta\nu)\log(V/V_0) = k\log\left[(V/V_0)^{\frac{E}{k\beta\nu}}\right]. \tag{7}$$

Now let us consider an ideal gas consisting of *n* point-like particles which move independently from each other and occupy uniformly the volume $V_0$, and let them have the total entropy $S_0$. The probability that a particle at some intsant of time can be found in a part of volume *V* is, of course $V/V_0$. The probability of the event that all the *n* particles occupy the volume *V* is then $w=(V/V_0)^n$, because the particles are assumed to be independent. According to Boltzmann's principle, the difference of the entropies belonging to these two states of the particle system equals $S-S_0=k\log w$,

$$S - S_0 = k \log w = k \log[(V/V_0)^n]. \tag{8}$$

By comparing Eqs. (7) and (8) we can conclude that if the ideal gas of total energy $E$ consist of $n$ identical and independent particles of energy $k\beta v = R\beta v/N = hv$, that is $E = nhv$, then the entropy difference calculated on the basis of Wien's formula coincides with the entropy difference coming out from the Boltzmann principle. (Here $R$ and $N$ denote the universal gas constant and the Avogadro number, respectively, and $R/N=k$.) With Einstein's words on page 143 [3][2] : "*Monochromatic radiation of small density (in the range of the validity of Wien's radiation formula) behaves so from the point of view of heat theory as if it consisted of independent energy quanta of size $R\beta v/N$.*" Already in the introduction of the paper Einstein makes even a more "revolutionary" statement on page 133 [3][3] : "*According to the assumption to be kept in eye here, by spreading from a point in the outgoing light rays the energy is not distributed continuously to larger and larger spatial regions, but these rays consist of a finite number of energy quanta localized in spatial points which move without falling apart, and they can be absorbed or created only as a whole.*" We emphasize, however, that *this more general statement, on the other hand, is not supported by a mathematical derivation*. It is interesting to note that in fact we encounter first here with the ideal light signals propagating in absolutely empty space, as was used by Einstein in his paper on the foundation of special relativity appeared four months later [13]. The word "photon" was introduced later, in 1926 [14], and then on its usage has gradually been widely spread. As an application of his hypothesis of light quanta, Einstein gave a natural explanation of the Stokes phenomena appearing in photoluminescence, the accurate results of Lenard's experiments on the photoelectric effect [15], and on Stark's observations on photoionization of gases. Though in the meantime the photoelectric effect has become the almost exclusive tool in photon counting and correlation experiments, we shall not discusse it any more, because it is out of the scope of the present paper.

### 3.2. *Einstein's fluctuation formula*

In 1909, by using Planck's formula Eq. (3) for the average energy of a mode of the radiation and the connection between entropy and thermodynamic probability, Einstein derived an expression for the fluctuation, the mean square deviation of the energy of the black-body radiation occupying a certain sub-volume in the *Hohlraum* [1]. This formula delivers a new support for his concept of light quanta. He writes on page 191 of Ref. [1][4] : "*We have seen that Planck's law of radiation can be derived on introducing the assumption that the energy of an oscillator of frequency $v$ can be built up only of quanta of magnitude $hv$. From this it is not necessary to take the assumption that the radiation, too, could be emitted and absorbed in quanta of such magnitude, because then here we would speak of a property of the emitting or absorbing material, respectively; however, the considerations 6 and 7 show that for the fluctuation of the spatial distribution of the radiation and of the radiation pressure, that kind of a formula comes out, as if the radiation consisted of quanta of the given magnitude. That*

---

[2] The original German text on page 143 of Ref. [3] : "Monochromatische Strahlung von geringer Dichte (innerhalb des Gültigkeitsbereichs der Wienschen Strahlungsformel) verhält sich in wärmetheoretischer Beziehung so, wie wenn sie aus voneinander unabhängigen Energiequanten von der Größe *R*β*v*/*N* bestünde."

[3] The original German text on page 133 of Ref. [3] : "Nach der hier ins Auge zu fassenden Annahme ist bei Ausbreitung eines von einem Punkte ausgehenden Lichtstrahles die Energie nicht kontinuierlich auf größer und größer werdende Räume verteilt, sondern es besteht dieselbe aus einer endlichen Zahl von in Raumpunkten lokalisierten Energiequanten, welche sich bewegen, ohne sich zu teilen und nur als Ganze absorbiert und erzeugt werden können."

[4] The original German text on page 191 of Ref. [1] : "Wir haben gesehen, daß das *Plancksche Strahlungsgesetz* sich begreifen läßt unter Heranziehung der Annahme, das Oscillatorenergie von der frequenz $v$ nur auftreten kann in Quanten von der Größe *h v*. Es genügt nach dem Vorigen nicht die Annahme, daß Strahlung nur in Quanten von dieser Größe *emittiert und absorbiert* werden könne, das es sich also lediglich um eine Eigenschaft der emittierenden bzw. absorbierenden Materie handle; die Betrachtungen 6 und 7 zeigen, daß auch die Schwankungen in der räumlichen Verteilung der Strahlung und diejenigen des Strahlungsdruckes derart erfolgen, wie wenn die Strahlung aus Quanten von der angegebenen Größe bestünden. Es kann nun zwar nicht behauptet werden, daß die Quantentheorie aus dem Planckschen Strahlungsgesetz als *Konsequenz* folge, und daß andere Interpretationen ausgeschlossen seien. Man kann aber wohl behaupten, daß die Quantentheorie die einfachste Interpretation der *Plancksche Formel* liefert."

*could not be asserted that the quantum theory would derive as a consequence of Planck's law of radiation and other interpretations could be excluded. However, one can safely state that the simplest interpretation of Planck's formula is given by the quantum theory."*

Let us first summarize Einstein's original derivation. Take two thermodynamically communicating spatial regions of volumes $V$ and $\upsilon$ enclosed by reflecting walls and filled out with thermal radiation in the frequency range ($\nu$, $\nu+d\nu$). When $H$ and $\eta$ are the instantaneous energies of the radiation in volume $V$ and $\upsilon$, respectively, then after a while the proportionality $H_0 : \eta_0 = V : \upsilon$ must hold within a good approximation, where $H_0$ and $\eta_0$ average values. At an arbitrary time $\eta$ will deviate from $\eta_0$ according to a statistical law determined by the Planck-Boltzmann relation $S = k \log W$, that is $dW = const \times exp(S/k) d\eta$. Now let us expand the entropy $S = \Sigma + \sigma$ into powers of the random deviation $\varepsilon$ which is defined by the equation $\eta = \eta_0 + \varepsilon$. By assuming that the deviations are small we keep terms only up to the second power, hence

$$S = const + \frac{1}{2}\left[\frac{d^2\sigma}{d\eta^2}\right]_0 \varepsilon^2 \quad \text{and} \quad dW = Const \times \exp\left\{-\frac{1}{2k}\left|\frac{d^2\sigma}{d\eta^2}\right|_0 \varepsilon^2\right\} d\varepsilon. \tag{9}$$

In deriving Eq. (9) we have tacitly assumed that to the mean value $\eta_0$ of the energy in the smaller volume $\upsilon$ belongs the maximum value of the entropy $\Sigma_0 + \sigma_0$, so the first derivative $[d(\Sigma + \sigma)/d\eta]_0 = 0$ vanishes, and the second derivative is negative. According to Eq. (9) the probability density $dW/d\varepsilon$ of the deviation is a gaussian distribution, hence

$$\Delta\eta^2 \equiv \overline{(\eta-\eta_0)^2} = \overline{\varepsilon^2} = \left\{\frac{1}{k}\left|\frac{d^2\sigma}{d\eta^2}\right|_0\right\}^{-1} = h\nu \cdot \eta_0 + \frac{c^3}{8\pi\nu^2 d\nu} \cdot \frac{\eta_0^2}{\upsilon}. \tag{10}$$

The last equation of Eq. (10) has been obtained by expressing from Planck's formula Eq. (3) $1/T = (k/h\nu)\log(1+h\nu m_\nu/\eta) = d\sigma/d\eta$, where $\eta = m_\nu U_1$ with $m_\nu = \upsilon \cdot (8\pi\nu^2/c^3) \cdot d\nu$ being the number of modes in the frequency interval ($\nu$, $\nu+d\nu$) in volume $\upsilon$. The second derivative of $\sigma$ is expressed as $d^2\sigma/d\eta^2 = -k\, m_\nu /(h\nu\, \eta\, m_\nu + \eta^2)$. Equation (10) is Einstein's famous fluctuation formula in its original form, which is the first mathematically correct quantum formula showing the "wave-particle duality" in case of the black-body radiation. We shall discuss its physical content soon. At the moment it is enough to state that the first term in Eq. (10) describes "particle-like fluctuations" and the second term describes "wave-like fluctuations" coming from random interferences of the Maxwell field. It is interesting to note that the latter equation for the second derivative of $\sigma$ is completely equivalent to Planck's original interpolation formula [10] $d^2\sigma/d\eta^2 = (1/m_\nu) \cdot d^2 S_1/dU_1^2 = (1/m_\nu) \cdot a/U_1(b+U_1)$ which was the first to lead to the correct spectral distribution presented by Planck on 19 October 1900. By leaving out the second term in the denominator we arrive at Wien's formula for $\eta/\upsilon d\nu = u_\nu \equiv \rho$ of Eq. (5), which was used by Einstein for the introduction of the photon concept. Planck's "fortunate interpolation" ment just to introduce the second, quadratic term $U_1^2$ in the denominator. Without this term the wave-like fluctuation would be missing on the right hand side of Eq. (10). It is well possible that Einstein had got the idea of this derivation by noticing the impotance of the quadratic term in Planck's interpolation formula, because he introduced his above analysis on page 188 of Ref. [1] with the words[5] : *"I have already completed a consideration of similar sort in an earlier work in which I first presented the theory of light quanta, in order to obtain certain statistical properties of heat radiation closed in a Hohlraum. That time I started in the limiting range (for small values of ν/T) where*

---

[5] The original German text on page 188 of Ref. [1] : "Eine Betrachtung der angedeuteten Art zur Ermittlung gewisser statistischer Eigenschaften von in einen Hohlraum eingeschlossener Wärmestrahlung habe ich bereits in einer früheren Arbeit[1]) durchgeführt, in der ich die Teorie der Lichtquanten zuerst darlegte. Da ich aber damals von der nur in der Grenze (für kleine Werte von ν/T) gültigen Wienschen Strahlungsformel ausging, will ich hier eine ähnliche Betrachtung angeben, welche eine einfache Deutung der Planckschen Strahlungsformel liefert."

*Wien's radiation formula is valid, but here I will give a similar consideration which delivers a simple meaning of Planck's radiation formula."* We note that in this quotation in the parenthesis the word "small" should be replaced by "large", as is already displayed correctly a couple of pages later in Einstein's paper.

The fluctuation formula for a system of oscillators was derived by Laue [16] by using the Planck-Bose distribution Eq. (4). By a simple calculation we obtain

$$\overline{n^2} = \overline{n} + 2\overline{n}^2, \quad \text{hence} \quad \overline{\Delta n^2} = \overline{n^2} - \overline{n}^2 = \overline{n} + \overline{n}^2,$$

$$\Delta E_\nu^2 = M_\nu (h\nu)^2 \overline{\Delta n^2} = h\nu \overline{E}_\nu + \overline{E}_\nu^2 / M_\nu, \text{ where } \overline{E}_\nu = M_\nu h\nu \cdot \overline{n} \tag{11}$$

If we identify $M_\nu$ with the degrees of freedom of the radiation field (with the number of modes) then the physical content of Eq. (11) is the same as that of Einstein's formula.

We can choose still another shorter way for the derivation of Eq. (10) based on a general relation of statistical mechanics [17]. Let $Z(\beta)$, with $\beta \equiv 1/kT$, be the canonical partition function (sum of states) of a system being in thermal equilibrium, $Z(\beta) = \int dE \Omega(E) e^{-\beta E}$, then $<E> = (1/Z) \int dE \Omega(E) E e^{-\beta E} = (1/Z)(-\partial Z/\partial \beta) = Z'/Z$, where the prime denotes derivative with respect to $\beta$. Here $\Omega(E)$ is the density of states, that is $\Omega(E)dE$ gives the number of states belonging to the interval $(E, E+dE)$. Elementary calculation shows that $\partial <E>/\partial \beta = -(Z''/Z)+(Z'/Z)^2$, so if we know the the functional dependence of the average energy on the absolute temperature, then the squared deviation (variance or fluctuation) of the energy can be calculated in complete generality with the help of the simple formula (we use throughout both the bracket $<>$ in the text and the upper dash in numbered formulae to denote mean values)

$$\Delta E^2 \equiv \overline{(E-\overline{E})^2} = \overline{E^2} - \overline{E}^2 = kT^2 \frac{\partial \overline{E}}{\partial T}. \tag{12}$$

Now let us apply this result for one mode with the average energy $U_1$ given by Planck's formula (2). We obtain

$$\Delta U_1^2 \equiv (\Delta E_{\nu 1})^2 = (h\nu)^2 \left[ \frac{1}{e^{h\nu/kT}-1} + \frac{1}{(e^{h\nu/kT}-1)^2} \right], \text{ and} \tag{13}$$

$$(\Delta E_{\nu 1})^2 = h\nu \overline{E}_{\nu 1} + \overline{E}_{\nu 1}^2. \tag{14}$$

If we multiply this with the number of modes $M_\nu = V \cdot (8\pi \nu^2/c^3) \cdot d\nu$ in volume $V$ in the frequency interval of width $d\nu$, then we arrive at the fluctuation formula equivalent to that of Einstein's, Eq. (11). Had we used Wien's limit formula, we would have only got the first term. On the other hand, in the Rayleigh-Jeans limit only the second term comes out.

### 3.3. *The wave-particle duality in the black-body radiation*

The two terms in Einstein's fluctuation formula, Eq. (11), or equivalently in Eqs. (13) and (14), have a very clear physical meaning as we show in the followings.

In order to give a physical interpretation of the first term in Eq. (11) let us consider a part $V$ of a *Hohlraum* of volume $V_0$ and assume that the average number of photons in $V$ is given by $<N> = (N_0/V_0)V$ in a certain frequency interval $(\nu, \nu+d\nu)$, where $N_0$ is the total number of photons (assumed now to be point-like particles) in this spectral range. The actual number $N$ of the photons in $V$ varies by chance from one instant to the other, and to a good approximation this random variable satisfies a Poisson distribution, as can be shown in the following way. Due to the assumed average homogenity and independence of the individual photons, the probability of finding exactly $N$ photons in $V$ is given by the binomial distrtibution $w(N) = [N_0!/N!(N_0-N)!](V/V_0)^N[1-(V/V_0)]^{N_0-N}$, because $N$ photons can be selected from the total number of $N_0$ in a number of $N_0!/N!(N_0-N)!$ different ways, each with the geometrical probability $V/V_0$. The probability that the remaining others of number $N_0 - N$ do *not* get into $V$ is just $[1-(V/V_0)]^{N_0-N}$, because of the assumed independence of the particles. If we take the limits $N_0 \to \infty$ and $V_0 \to \infty$, in such a way that $N_0(V/V_0) \equiv \rho \cdot V \equiv \langle N \rangle$ remain finite, where $\rho$ denotes the photon density, then the above binomial distribution goes over to a *Poisson distribution* [18], [19]. This can be checked by using Stirling's formula $N! \to (N/e)^N$, hence

$$w(N) = \lambda^N e^{-\lambda} / N!, \quad \lambda = \overline{N}, \quad \Delta N^2 \equiv \overline{N^2} - \overline{N}^2 = \overline{N}, \tag{15}$$

$$\Delta E^2 = (h\nu)^2 (\Delta N)^2 = h\nu \overline{E}. \tag{16}$$

In Eq. (15) we have displayed the mean square deviation $\Delta N^2$ of the number of particles in volume $V$, and in Eq. (16) we have given the mean square deviation $\Delta E^2$ of the corresponding energy, in the case when all the particles have the energy $h\nu$ (within the spectral range ($\nu$, $\nu+d\nu$)). The term on the right hand side of the latter equation looks exactly like the first term in the formula given by Eq. (11). So, according to the above consideration we may state that the first term in Einstein's fluctuation formula Eq. (10) accounts for *particle-like fluctuations*. For visible light ($h\nu \sim 1eV$) at room temperature ($kT \sim 0.025eV$) we have $h\nu/kT \sim 40 \gg 1$, that is Wien's formula is a good approximation, and particle-like fluctuations dominate. At the surface of the Sun at $\sim 6000\ K$ the ratio $h\nu/kT \sim 2$, so the wave-like fluctuations still do not overtake, which can only happen in the visible well above $\sim 12000\ K$. Of course, the relative fluctuations depend on the number of modes in the sub-volume under discussion. It is interesting to note that the particle-like fluctuations of light can also be observed by human eye as was shown for instance by Wawilow [20] in a series of experiments starting in the thirties of the last century.

We also mention here de Broglie's remark [6] made in 1922 concerning Eqs. (13) and (14). The sum of the two terms (13) describing the fluctuation of a particular mode can be expanded into an infinite sum,

$$(\Delta E_{\nu 1})^2 = h\nu \overline{E}_{\nu 1}^{(1)} + 2h\nu \overline{E}_{\nu 1}^{(2)} + \ldots = \sum_{m=1}^{\infty} m h\nu \overline{E}_{\nu 1}^{(m)}, \quad \overline{E}_{\nu 1}^{(m)} \equiv h\nu e^{-mh\nu/kT}, \tag{17}$$

where the last definition refers to the kind of Wien's distributions corresponding to independent classical ideal gases consisting of "*photo-molecules*" or "*photo-multiplets*" of energies $mh\nu$ with $m = 1, 2, 3, \ldots$ . As a consequence, a component of the heat radiation in a mode can be considered – at least from the point of view of its energy and statistics – as a mixture of infinitely many noninteracting ideal gases following the Boltzmann statistics. In this way the complete fluctuation has been decomposed into a sum of purely particle-like fluctuations. If one keeps only the first term in Eq. (17) then one gets back to Einstein's original "single photons" corresponding to Wien's approximate radiation formula. In order to get the wave-term, one has to include all the higher terms. We shall see in Section 6. that this completely corpuscular interpretation of the Planck formula can be exactly founded in all details by using classical statistics, thermodynamics and probability theory.

Concerning the wave nature of the heat radiation, an analogous expression to the second term in Einstein's fluctuation formula, Eq. (10), can be derived from the stochastic wave character of the radiation in the following sense. Let the electric field of a component of the radiation be expressed as $a_\nu(t) = a_c \cos\omega t + a_s \sin\omega t = a_\nu \cos(\omega t - \theta_\nu)$, where $\omega \equiv 2\pi\nu$ is the circular frequency, and $a_\nu \equiv \sqrt{(a_c^2 + a_s^2)}$ and $\theta_\nu \equiv \arcsin(a_s / a_\nu)$ are the amplitude and phase of the oscillation, respectively. We can imagine the field of the chaotic radiation as a superposition of a very large number of small independent contributions, for example
$a_c = a_{c1} + a_{c2} + a_{c3} + \ldots$, where the summands are random amplitudes of *arbitrary distributions* stemming from a large number of radiators of the surroundings. Let $a_{cn} / a_{\nu 1} = (1/a_{\nu 1}\sqrt{n})(a_{c1,n} + a_{c2,n} + \ldots + a_{cn,n})$, where $a_{\nu 1} > 0$ denotes the square root of the arithmetic mean of the (finite) variances of the components. The building up of the part of the field proportional with $a_c$ can be imagined in such a way that, departing from 0, it undergoes a random walk on the real axis. It it is clear that the expectation value of $a_c$ is zero, if the small components do not have a drift. By assuming the same for the sine component, we see that the formation of the complete complex amplitude $a_n = a_{cn} + i a_{sn} = |a_n| e^{i\theta_n}$ can be imagined as a result of a random walk on the complex plane starting from the origin, whose real and imaginary displacements are independent. In the limit $n \to \infty$, according to the Central Limit Theorem [18] [19], the distribution functions of $a_{cn}$ and $a_{sn}$ go over into the normal distribution, $P(a_{cn}/a_{\nu 1} < x) \to \Phi(x)$ and $P(a_{sn}/a_{\nu 1} < x) \to \Phi(x)$, where $\Phi$ denotes the Gauss error function, hence the probability *densities* are Gauss functions,

$$f_c(a_c) = \left(1/\sqrt{2\pi} a_{\nu 1}\right) \exp(-a_c^2 / 2 a_{\nu 1}^2), \tag{18a}$$

$$f_s(a_s) = \left(1/\sqrt{2\pi} a_{\nu 1}\right) \exp(-a_s^2 / 2 a_{\nu 1}^2). \tag{18b}$$

The stochastic average of the energy density of the mode, $[< a_\nu^2(t) >/8\pi]d\nu = [a_{\nu 1}^2/8\pi]d\nu$ has to be equal to the product of the mode density $(8\pi\nu^2/c^3) \cdot d\nu$ and the mean energy $<E_{\nu 1}>$ of the mode considered. From this requirement we obtain $[a_{\nu 1}^2/8\pi] = (8\pi\nu^2/c^3)<E_{\nu 1}> = Z_\nu <E_{\nu 1}>$. According to the statistical definition of the entropy we have

$$S_{\nu 1} = -k \int_{-\infty}^{\infty} da_c \int_{-\infty}^{\infty} da_s \, f_c(a_c) f_s(a_s) \{\log[\alpha f_c(a_c)] + \log[\alpha f_s(a_s)]\}, \tag{19a}$$

$$S_{\nu 1} = k \log(\overline{E}_{\nu 1} / E_0), \quad E_0 \equiv \alpha^2 /(2\pi e)(8\pi Z_\nu), \tag{19b}$$

where the factor $\alpha$ in Eq. (19a) has been introduced merely to make the arguments of the logarithms dimensionless. By using the result Eq. (19b), from the basic thermodynamical equation $\partial S_{\nu 1}/\partial E_{\nu 1} = 1/T$ we obtain

$$\partial S_{\nu 1}/\partial \overline{E}_{\nu 1} = 1/T \;\to\; \overline{E}_{\nu 1} = kT \;\to\; \partial^2 S_{\nu 1}/\partial \overline{E}_{\nu 1}^2 = -k/\overline{E}_{\nu 1}^2, \tag{19c}$$

which is just expressing the equipartition of energy for the radiation modes, from which the Rayleigh-Jeans law follows, $u_\nu = Z_\nu kT$. Because the joint distribution $f_c \cdot f_s$ does not depend on the phase $\theta$, the energy distribution of a mode is a simple exponential distribution of the random variable $E_1 = (a_c^2 + a_s^2)/8\pi Z_\nu$,

$$f(E_1) = (1/\overline{E}_1) e^{-E_1/\overline{E}_1}, \quad \text{hence} \quad \Delta E_1^2 \equiv \overline{E_1^2} - \overline{E}_1^2 = 2\overline{E}_1^2 - \overline{E}_1^2 = \overline{E}_1^2. \tag{20}$$

From Eq. (20) we can obtain an expression for the wave fluctuation, completely analogous to the second term in Einstein's fluctuation formula, Eqs. (11) or (10),

$$(\Delta E_\nu)^2 = \upsilon \cdot Z_\nu d\nu \cdot \Delta E_1^2 = m_\nu \cdot \Delta E_1^2 = \overline{E}_\nu^2 / m_\nu, \tag{21}$$

where $E_\nu = m_\nu E_1$ is the total energy in the spectral range $(\nu, \nu + d\nu)$.

We would like to emphasize here that, although both the purely particle-like fluctuation, Eq. (16), and the purely wave-like fluctuation, Eq. (21), have the *same form* as the corresponding terms in Einstein's fluctuation formula, Eqs. (11) or (10), nevertheless neither of them coincides *numerically* with the corresponding terms. The numerical agreement can be secured only in the corresponding limiting cases, namely in the Wien limit ($h\nu/kT \gg 1$), and in the Rayleigh-Jeans limit ($h\nu/kT$) $\ll 1$, respectively. This is because the average energies differ from the approximate values in the general case. The same is also true for the fluctuation expression, Eq. (17) coming from the statistically independent photo-multiplets. Similarly, we cannot state that for the wave-like fluctuations only the higher terms are responsible in general.

In addition, we note that the Planck-Bose distribution, Eq. (4) has first been written down by Planck [22], and used later by him [26] and by von Laue [16] after the introduction of his "second theory" [21-24] and his "third theory" [25], in which he derived the zero-point energy $h\nu/2$ for the oscillator. A very clear presentation of the wave-like fluctuations can be found in Planck [27-28]. He was the first to derive in [27] the exponential distribution given in Eq. (20) from the wave theory by using an alternative method to ours presented here. We also mention that von Laue [16] used the Planck-Bose distribution in 1915 to calculate the energy fluctuations of a solid body consisting of harmonic oscillators, which was first discussed by Einstein [29] at the Solvay Meeting in 1911.

**3.4. *Fluctuation of the radiation pressure experienced on a mirror in a Hohlraum***
In his first paper on the fluctuations of the black-body radiation [1] Einstein devoted a section to the analysis of the brownian motion of a plane mirror moving along a straight line parallel to its normal in a *Hohlraum* surrounded completely by matter of absolute temperature *T*. He argues that if the mirror is being in motion, then its front side reflects more radiation than the rare side, so there will appear a friction-like force acting on the mirror. Consequently, the momentum of the mirror will again and again change resulting in a fluctuation of the radiation pressure. By determining these fluctuations one can draw conclusions on the constitution of the radiation and, moreover, on the nature of the elementary processes of reflections taking place at the mirror.

Let us assume that the mirror has a velocity $u$ at some instant of time $t$, and during a small time interval $\tau$ this velocity is decreased by $Pu\tau/m$, where $P$ represent the retarding force per unit velocity of the mirror of mass $m$. The velocity of the mirror at the time instant $t + \tau$ is $u - Pu\tau/m + \delta$, where $\delta$ denotes the change in the velocity during $\tau$ caused by the random fluctuations of the radiation pressure. We require that $u$ does not change – at least on average – during the small time interval $\tau$, i.e.

$$\overline{(u - Pu\tau/m + \delta)^2} = \overline{u^2} \text{ , consequently } \overline{\delta^2} = (2P\tau/m)\overline{u^2} \text{ ,} \quad (22)$$

where we have assumed that $<u\delta>=0$, and canceled the term quadratic in $\tau$. (The stochastic average is denoted by either the bracket $<\ >$ or with the upper dash in the numbered formulae.) The average kinetic energy of the mirror follows from the equipartition theorem, $m<u^2> = kT$, so the mean square of the random momentum $m\delta = \Delta$ satifies the equation

$$\frac{\overline{\Delta^2}}{\tau} = 2kTP = 2kT \frac{3}{2c}\left[\rho - \frac{1}{3}v\frac{d\rho}{dv}\right]fdv \text{ ,} \quad (23)$$

where $\rho$ is the spectral density of the radiation and $f$ is the area of the mirror. In obtaining Eq. (23) Einstein assumed that the mirror is reflecting in the spectral range between $v$ and $v + dv$, and for other colors it is completely transparent. The expression for the friction coefficient $P$ given in the second equation in Eq. (23) was calculated by Einstein and Hopf [30], [31]. Later a similar expression was used by Einstein and Stern [32] in their derivation of Planck's formula without assuming any discontinuity, but a zero-point oscillation. By using Planck's law, Eq. (3), for the spectral density $\rho = u$ we obtain from Eq. (23) the formula for the momentum fluctuation,

$$c^2\overline{\Delta^2} = \left[hv\rho + \frac{c^3\rho^2}{8\pi v^2}\right]fc\tau \cdot dv \text{ .} \quad (24)$$

The similarity of the two fluctuation formulae for the energy (10) and for the momentum (24) is immediately seen, in particular when we use an alternative form of (10)

$$\overline{\varepsilon^2} = \left[hv\rho + \frac{c^3\rho^2}{8\pi v^2}\right]v \cdot dv \text{ .} \quad (25)$$

If we identify the volume $v$ with $fc\tau$, then we conclude $<\varepsilon^2> = c^2 <\Delta^2>$, which shows the well known relation between the photon energy and momentum $p = \varepsilon/c = hv/c$. It should be noticed that in each formulae given by Eqs. (24) and (25) the particle-like and the wave-like fluctuations simply add, as if they had two independent causes [33]. The first term for visible light can be much larger than the second one. Take, for instance, Einstein's numerical example, namely $\lambda \sim 0.5\mu$ and $T \sim 1700$ K, then the particle-term is about $7\times10^7$ times larger than the wave-term. The other remarkable result is that the momentum fluctuation, Eq. (24) is proportional to the area $f$ of the mirror. From this circumstance it follows that the pressure fluctuations coming from neighbouring parts of the plate (whose linear dimensions are large in comparison with the wavelength of the reflected radiation) are independent from each other. This suggests the picture again that in the Wien limit the radiation consists of spatially localized complexes of energy $h\nu$.

We close the present subsection with Einstein's words summarizing his view on the structure of the radiation based on the above analysis[6] : " *Nevertheless, for the time being, the most natural notion seems to me, that the appearance of the electromagnetic fields of light would also be attached to singular points, like in the case of electrostatic fields according to the electron theory. It is not excluded that in such a theory the energy of the electromagnetic field could be viewed as localized in these singularities, like in the old action-at-a-distance theory. I think of such singular points surrounded by force fields, which, in essence are of a character of plane waves, whose amplitudes decrease by the distance from the singular points. If there are many such singularities in a region, then their force fields will be on top of each other, and this assembly will form an undulatory force field, which, perhaps could hardly be distinguished from an undulatory field in the sense of the present theory of light. Needless to say, such a picture is of no value until it leads to an exact theory. With the help of it I merely wanted to illustrate in short that each of the structural properties (the undulatory structure and the quantal structure) which both show up according to Planck's formula, should not be viewed as incompatible to each other.*"

## 4. Bose Distribution from combinatorial Analysis and Fluctuations from Wave Interference

### 4.1. *Photon distributions from combinatorial analysis*

The Planck-Bose distribution Eq. (4) has already been presented by Hendrik Antoon Lorentz in 1910 during the famous Göttingen Lorentzwoche in his sixth lecture [34]. A couple of months later L. Natanson published a very clear combinatorial analysis of the the possible distributions of "energy elements" among "receptacles of energy" [35]. He used essentially Boltzmann's method of energy discretization, which had already been publised in 1877. We shall mostly follow his line of thought in the present subsection because one of his important results was rediscovered thirteen years later by S. N. Bose in his famous derivation of Planck's law of black-body radiation [36]. The energy elements $\varepsilon$ may represent both Planck's energy quanta or Einstein's photons and, accordingly, the receptacles can be imagined as material resonators, Bose's phase-space cells, or – which is the same – the normal modes of the radiation field in a *Hohlraum*, or even Laues's "Strahlenbündel" (bundles of rays of radiation). *The N independent receptacles can contain $0\cdot\varepsilon$, $1\cdot\varepsilon$, $2\cdot\varepsilon$, … , $p\cdot\varepsilon$ energy elements* (*where p can be very large, in effect, it may go to infinity*), *and we denote by $N_0$, $N_1$, $N_2$, …, $N_i$, … the number of receptacles which contain no quanta, one quantum, two quanta, etc*. If $E$ is the total energy contained in the system, than there are altogether $n = E/\varepsilon$

---

[6] The original German text on pages 824-825 of Ref. [33] : "Immerhin erscheint mir vorderhand die Auffassung die natürlichste, daß das Auftreten der elektromagnetischen Felder des Lichtes ebenso an singulären Punkte gebunden sei wie das Auftreten elektrostatischer Felder nach der Elektronentheorie. Es ist nicht ausgeschlossen, daß in einer solchen Theorie die ganze Energie des elektromagnetischen Feldes als in diesen Singularitäten lokalisiert angesehen werden könnte, ganz wie bei der alten Fernwirkungstheorie. Ich denke mir etwa jeden solchen singulären Punkt von einem Kraftfeld umgeben, das im wesentlichen den Charakter einer ebenen Welle besitzt, und dessen Amplitude mit der Entfernung vom singulären Punkte abnimmt. Sind solcher Singularitäten viele in Abständen vorhanden, so werden die Kraftfelder sich übereinanderlagern und in ihrer Gesamtheit ein undulatorisches Kraftfeld ergeben, das sich von einem undulatorischen Felde im Sinne der gegenwärtigen Lichttheorie vielleicht nur wenig unterscheidet. Daß einem derartigen Bilde , solange dasselbe nicht zu einer exakten Theorie führt, kein Wert beizumessen ist, braucht wohl nicht besonders hervorgehoben zu werden. Ich wollte durch dasselbe nur kurz veranschaulichen, das die beiden Struktureigenschaften (Undulationsstruktur und Quantenstruktur), welche gemäß der Planckschen Formel beide der Strahlung zukommen sollen, nicht als miteinander unvereinbar anzusehen sind. "

elements or quanta at disposal. It is clear that the following constraint relations have to be satisfied

$$\sum_{i=0}^{p} N_i = N \quad \text{and} \quad \sum_{i=0}^{p} i \cdot N_i = n. \qquad (26)$$

We call a distribution in which the above constraint relations are satisfied a *mode of distribution* ("Verteilungsart"), and we will symbolize it by the scheme

$$\begin{Bmatrix} 0\varepsilon & 1\varepsilon & 2\varepsilon & \dots & p\varepsilon \\ N_0 & N_1 & N_2 & \dots & N_p \end{Bmatrix}. \qquad (27)$$

In order to specify a mode of distribution we merely have to know how many receptacles contain zero, one, two, etc. elements, but we do not keep track of neither the individual elements nor the individual receptacles.

The conditions are considerably different when we suppose that each receptacle can be identified, in other word denumerated. This is the case for instance with the normal modes of a *Hohlraum*. When the "first" receptacle contains $n_1$ elements, the second $n_2$ elements, and so on, then we have a *mode of collocation* ("Anordnungsart"), which is symbolized this way

$$\begin{Bmatrix} I & II & III & \dots & N \\ n_1 & n_2 & n_3 & \dots & n_N \end{Bmatrix}. \qquad (28)$$

The numbers $n_1, n_2, n_3, \dots$ can take on the values 0, 1, 2, … , and we know that in the sequence $n_1, n_2, n_3, \dots, n_N$ the number 0 $N_0$ – times, the number 1 $N_1$ – times etc. will appear. The number of collocations Eq. (28), $A(N_0, N_1, N_2, \dots, N_p)$, which can be obtained from the mode of distribution Eq. (27), satisfy the relation

$$A(N_0, N_1, \dots, N_p) \prod_{i=0}^{i=p} N_i! = N! . \qquad (29)$$

In the case of collocations we distinguish the receptacles but the quanta are considered indistinguishable. This is just the essence of the Bose statistics.

The situation fundamentally changes if we keep track of the individual quanta. Let us name the $n_1$ quanta $a, b, c, \dots$ , the $n_2$ quanta $f, g, h, \dots$, and so on. This way we obtain a *mode of association* ("Zuordnungsart")

$$\begin{Bmatrix} I & II & \dots \\ a,b,c & f,g,h & \dots \end{Bmatrix}. \qquad (30)$$

The number of associations Eq. (30), $B(N_0, N_1, N_2, \dots, N_p)$, which can be obtained from the mode of distribution Eq. (27), satisfiy the relation

$$B(N_0, N_1, \dots, N_p) \prod_{i=0}^{i=p} (i!)^{N_i} = n!. \qquad (31)$$

Since $B$ depends only on the numbers $N_0, N_1, N_2, \dots, N_p$, the number of associations are the same for any collocations belonging to the distribution Eq. (27). Thus, all the distribution for which $A$ collocations of the energy are possible, will correspond to $A \cdot B$ associations. Since the quanta are distinguishable in this case, we are dealing with the Maxwell-Boltzmann statistics. It can be shown that the sum of the number of all collocations Eq. (29) compatible with the subsidiary conditions Eq. (26) are given as

$$\sum_{N_0, N_1, \dots, N_p} A(N_0, N_1, \dots, N_p) = \frac{(N+n-1)!}{(N-1)! n!} , \qquad (32)$$

and the sum of the number of associations belonging to all the allocation

$$\sum_{N_0, N_1, \dots, N_p} A(N_0, N_1, \dots, N_p) \cdot B(N_0, N_1, \dots, N_p) = N^n . \qquad (33)$$

The corresponding probabilities $P$ and $Q$ of the allocations and of the associations are

$$P(N_0, N_1, ..., N_p) = A / \sum A \quad \text{and} \quad Q(N_0, N_1, ..., N_p) = AB / \sum AB. \quad (34)$$

The equilibrium distributions of the collocations (undistinguishable quanta) and of the associations (distinguishable quanta) can be obtained by the well-known Lagrange multiplier technique by varying the entropies $S_c = k\log P$ and $S_a = k\log Q$, respectively [35]

$$N_i^{Bose} = N\left(1 - e^{-\varepsilon/kT}\right)e^{-i\varepsilon/kT}, \quad \text{and} \quad N_i^{Maxwell} = (N/kT)\varepsilon e^{-i\varepsilon/kT}. \quad (35)$$

The first expression of Eq. (35) is equivalent to the Planck-Bose distribution Eq. (4) of subsection 2.2, which leads in the case of the black-body radiation to the form of Einstein's fluctuation formula (11), obtained by Laue in 1915 [16]. In the continuum limit ($\varepsilon \to 0$) the second expression in Eq. (35) goes over to the first formula of Eq. (20) of subsection 3.3, which leads to the wave-like fluctuation.

*By closing the present subsection, we wish to emphasize that the very Bose distribution given in the first equation of Eq. (35) – where N, the number of energy receptacles, should be identified with the number of cells or modes $N=M\equiv V(8\pi v^2/c^3)dv$ in a Hohlraum – was found by Natanson by applying in fact Bose's statistics in 1911, that is, thirteen years earlier than Bose's well-known article appeared in 1924 [36]. Moreover, if we choose p = 1 in Eq. (27), so we take the largest possible number of elements occupying the receptacles to unity, then from Natanson's analysis we can deduce the Fermi distribution, too.*

### 4.2. *Fluctuations from wave interference*

Concerning the wave nature of the heat radiation we have already derived in subsection 3.3 an expression for the fluctuation, Eq. (21), which is an analogon of the second term in Einsteins fluctuation formula, Eq. (10) or Eq. (11). However, in that analysis we used stochastic ensembles to describe the Fourier components of the field amplitudes, and the interferences of the components have not been shown up explicitly. *The wave-like fluctuation can also be viewed as a result of an irregular "breeding" of the beat waves formed from the interfering Fourier components.* This point is quite important here, because Einstein's two-term fluctuation formula, Eq. (11), appears to be inconsistent with the one computed from the interference of the wave trains. Eq. (11) has been held to indicate a fundamental inconsistency between the the electromagnetic theory on the one hand, and the Planck law on the other [37]. As Jacobson remarked in 1927: "This conclusion has not been accepted by Laue[5] nor by Planck[6]. The fundamental point at issue is the statistical independence of the constituent wave trains in the fourier analysis of natural radiation. A proof of this independence was given by Einstein and Hopf[7] on the hypothesis that the radiation arose from an independent number of point sources: later in response to the objections of Laue a proof was given by Einstein[8] on somewhat different hypotheses. More recently, the question has been taken up again by Planck[6] who, starting from a simplified form of Eq. (1) applicable to a system of only one degree of freedom, which has been given by Laue (Eq. (14) in the present paper, S. V.) av[(E-$E_{av}$)$^2$]=hv·$E_{av}$+($E_{av}$)$^2$ shows that the required condition is satisfied if there exist phase relations of a certain type between the various harmonics in the fourier analysis of the vibration." [37]. In the followings we briefly summarize the basic points in Planck's analysis [28].

The Fourier series of the electric field of a stationary monochromatic natural radiation of one spatial degree of freedom in the time interval $0 < t < T$ can be written in the form

$$A(t) = \sum_n C_n \cos(2\pi nt/T + \vartheta_n), \quad C_n > 0, \quad |(n-n_0)/n_0| \ll 1, \quad (36)$$

where $T$ is a very large time value and the central frequency is given by $v = n_0/T$. The "slowly varying" energy $E$ of the radiation can be obtained by averaging $A^2(t)$ on a time interval being much larger than an oscillation period, but it is much smaller then the time scale of the fluctuations

$$E = K \cdot (a_0 / 2) + K \cdot \sum_m [a_m \cos(2\pi mt/T) - b_m \sin(2\pi mt/T)], \tag{37}$$

$$a_0 = \sum_n C_n^2 \tag{38}$$

$$a_m = \sum_n C_{m+n} C_n \cos(\vartheta_{m+n} - \vartheta_n), \quad b_m = \sum_n C_{m+n} C_n \sin(\vartheta_{m+n} - \vartheta_n), \tag{39}$$

where $m \ll n$. According to the latter inequivality the characteristic times of the fluctuations are much smaller than the oscillation frequencies $\sim 1/\nu$. In Eq. (37) we have introduced a suitably choosen constant $K$. The time average of $E$ calculated over the time interval $0 < t < T$ reads simply

$$\overline{E} = K a_0 / 2 = (K/2) \sum_n C_n^2, \tag{40}$$

and the fluctuation is given as

$$Q = \overline{(E - \overline{E})^2} = (K^2/2) \sum_m (a_m^2 + b_m^2) \equiv Q_1 + Q_2, \tag{41}$$

where we have split the fluctuation into two parts $Q_1$ and $Q_2$. The first term in Eq. (41) gives the purely wave-like fluctuation term

$$Q_1 = \left[ (K/2) \sum_n C_n \right]^2 = \overline{E}^2. \tag{42}$$

The second term in Eq. (41)

$$Q_2 = K^2 \sum_{m,q,n} C_{m+q+n} C_{m+n} C_{q+n} C_n \cos(\vartheta_{m+q+n} - \vartheta_{m+n} - \vartheta_{q+n} + \vartheta_n) \tag{43}$$

can be brought to a form $h\nu \langle E \rangle$ (i. e. to the form of the particle-term) if we require the following equality to hold

$$K \sum_{m,q,n} C_{m+q+n} C_{m+n} C_{q+n} C_n \cos(\vartheta_{m+q+n} - \vartheta_{m+n} - \vartheta_{q+n} + \vartheta_n) = (h\nu/2) \sum_n C_n^2. \tag{44}$$

In general it is assumed that there are no definite phase relations between the Fourier components, and then the triple sum averages out to zero, hence $Q_2 = 0$ in Eq. (41), that is, according to Eq. (42) we have only the wave-like fluctuation $Q_1 = \langle E \rangle^2$.

Planck has constructed a classical field satisfying the requirements of natural radiation which produces the particle-like fluctuation too, for which $Q_2 = h\nu \langle E \rangle$, and this way the complete fluctuation formula is recovered, at least formally. In order to see this let us take a system of a large number of sine waves with considerably different orders $n_1, n_2,..., n_P$ schwitched on at time instants $t_1, t_2,..., t_P$, and schwitched off after a common time interval of size $\tau$. The $n$-s are independent of the $t$-s and each of them are distributed irregularly such that the quasi-monochromaticity condition in Eq. (36) is satisfied, and $0 < t_1 < t_2 < ... < t_P < T$. Moreover, we require that

$$(T/n_0) \ll [(n_i - n_0)/n_0]\tau \ll \tau \ll T, \tag{45}$$

which means that the duration $\tau$ of the individual pulses is much larger then the oscillation period, but it is much smaller than the complete duration $T$. Now take the field

$$A = \sum_{i=1}^{P} C \sin[(2\pi n_i / T)(t - t_i)] \cdot f[t_i, t_i + \tau], \tag{46}$$

where $f[t_i, t_i+\tau]$ are envelop functions whose values are 1 inside the carrier interval $[t_i, t_i+\tau]$, and outside they are zero. After carrying out the time averaging, we have

$$\overline{E} = KC^2 P\tau / 2T, \quad Q = \overline{(E - \overline{E})^2} = (KC^2/2)\overline{E} + \overline{E}^2. \tag{47}$$

If we choose $KC^2/2 = h\nu$ in Eq. (47), then we recover (formally) the complete fluctuation expresssion which looks like Laue's expression Eq. (14). *This way, besides the wave-like fluctuation, we were able to derive the particle-like fluctuation from mere wave interference.*

*In order to obtain this result, we had to prescribe a definite phase relation given in Eq. (46) between the Fourier components of the wave train.* At the end of his paper Planck emphasized that, though interesting, this was far not a satisfactory result, since the correspondence had been proved only up to the second moments and only for one degree of freedom.

## 5. Fluctuation from Classical Randomness, Spontaneous Emission and Field Quantization

### 5.1. *The contributions by Ehrenfest and Smekal to the theory of fluctuations*

In 1925 Paul Ehrenfest [39] gave a detailed analysis concerning the energy fluctuations due to the superposition of eigenmodes of a *Hohlraum* whose energies were assumed to be classical random variables following the Planck-Bose distribution given in Eq. (4). According to his results the energy fluctuation is reduced when we take into account the superposition of such "quantized eigenmodes". He started his discussion by drawing attention to the following appearent contradiction: Einstein derived his formula for the fluctuation of the energy of black-body radiation in a sub-volume of the Hohlraum by using Planck' formula, and arrived at the result which cannot be reproduced by assuming pure wave interference (see the first term on the right hand side of Eq. (10)). On the other hand, the Planck formula itself can be derived by introducing the energy level statistics of *waves* in a Jeans cube (as was shown much earlier by Debye [40]). So, how is it? We start with the wave description, and through the Planck formula we derive a fluctuation term which does not match to the wave picture? As Ehrenfest stated, the answer had already been given by Ornstein and Zernike [41] in 1919, who – according to Ehrenfest – claimed that in case of wave interference the additivity of the entropies of the sub-volume elements in a Hohlraum is not satisfied, in contrast to the spirit of Einstein's original derivation. In fact, these authors had not used such an argument in their publication, as is clearly stated in [42].

Ehrenfest derived two new formulas for the relative fluctuations which may be considerably different from that of Einstein, Eq. (11), in particular when the sub-volume $\upsilon$, in which the fluctuation is considered, is much smaller than the whole volume $V$ of the *Hohlraum*. Since he discussed relative fluctuations, let us first give an alternative form of Einstein's original fluctuation formula, Eq. (11).

$$\frac{\overline{(E_\nu - \overline{E}_\nu)^2}}{\overline{E}_\nu^2} = \frac{1}{n \cdot m_\nu} + \frac{1}{m_\nu}, \text{ where} \qquad (48a)$$

$$\overline{n} \equiv \frac{1}{e^{h\nu/kT} - 1}, \quad \text{and} \quad m_\nu = \upsilon \cdot \frac{8\pi\nu^2}{c^3} d\nu \qquad (48b)$$

In Eq. (48b) we have introduced the number of modes $m_\nu$ in the frequency range ($\nu$, $\nu+d\nu$) in the sub-volume $\upsilon$. It is interesting to note that the two terms in Eq. (48a) are equal if the average photon occupation number is unity. In this case the photon energy $h\nu=kT\log 2$ just equals to the minimum energy amount necessary to transmit one bit of information according to Shannon.

In order to make the calculations simpler, Ehrenfest considered a simplified model, namely the transverse oscillations of a string with fixed end points in a plane. A very thorough discussion of this problem can be found for instance in the excellent book by Fetter and Walecka [43]. The wave equation of the perpendicular elongation $u(x,t)$ of the string of length $L$ satisfies the following wave equation and boundary conditions

$$\frac{1}{c^2}\frac{\partial^2 u(x,t)}{\partial t^2} = \frac{\partial^2 u(x,t)}{\partial x^2} \text{ , where } c^2 = \frac{\tau}{\sigma} \text{ , and} \quad (49a)$$

$$u(0,t) = u(L,t) = 0 . \quad (49b)$$

In Eq. (49a) $\tau$ and $\sigma$ are the (constant) tension and the linear mass density of the string, respectively, and $c$ is the propagation velocity of the transverse waves. The Bernoulli solutions of Eqs. (49a-b) are given by the Fourier series of the normal modes

$$u(x,t) = \sum_{n=1}^{\infty} C_n (2/\sigma L)^{1/2} \sin(nkx)\cos(n\omega t + \phi_n) , \quad (50a)$$

where
$$k = \pi/L , \quad \omega = ck . \quad (50b)$$

In Eq. (50a) $C_n$ and $\phi_n$ are the by now arbitrary amplitudes and phases of the oscillations of the spectral components. The total energy $H$ of the string is a constant of motion,

$$H = (1/2)\int_0^L dx[\sigma(\partial u/\partial t)^2 + \tau(\partial u/\partial x)^2] = (1/2)\sum_{n=1}^{\infty} \omega_n^2 C_n^2 , \quad (50c)$$

where $\omega_n = n\omega$. The energy $E(t)$ of a piece of length $l$ of the string is given by the expression

$$E(t) = (1/2)\int_{(l)} dx[\sigma(\partial u/\partial t)^2 + \tau(\partial u/\partial x)^2] . \quad (51)$$

Now let us consider the contributions to the energy $E(t)$ from a narrow spectral range corresponding to high harmonic indeces. This part of the energy will be denoted by $e(t)$. It can be shown by a straightforward calculation that this equals

$$e(t) = \sum_{n,m} B_n B_m \cos[(n-m)\omega t + \phi_n - \phi_m] I_{n,m} , \quad (52a)$$

where $\quad B_n = n\omega C_n / L^{1/2} , \text{ and } I_{n,m} = (1/2)\int_{(l)} dx \cos[(n-m)kx] . \quad (52b)$

According to Ehrenfest, in the double sum in Eq. (52a) only those terms are taken into account for which the following inequalities hold

$$n,m \gg 1, \ |n-m|/(n+m) \ll 1, \ \omega' < n\omega, m\omega < \omega' + d\omega' \quad (52c)$$

The physical meaning of the first condition is that we are dealing with oscillations whose wavelenths $\lambda_n = 2L/n$ are much smaller that the length of the string. The second and third inequality mean that we consider a narrow spectral range around a central frequency. In obtaining Eq. (52a) it has also been assumed that the size $l$ of the "sub-volume" is much larger than the wavelengths $\lambda_n$. Ehrenfest considered the amplitudes and phases in the double sum in Eq. (52a) as independent classical random variables, such that

$$\overline{\cos(\phi_n - \phi_m)} = \overline{\sin(\phi_n - \phi_m)} = 0 , \quad (53a)$$

$$\overline{B_n \cdot B_m} = \overline{B_n} \cdot \overline{B_m} , \ \overline{B_n^2} = \overline{B_m^2} \equiv \overline{B^2} , \ \overline{B_n^4} = \overline{B_m^4} \equiv \overline{B^4} . \quad (53b)$$

By taking the above two equations into account, the ensemble average of the energy $e(t)$ in Eq. (52a) becomes

$$e_0 \equiv \overline{e(t)} = \sum_n \overline{B_n^2} I_{n,n} = \frac{l}{2}\overline{B^2} Z , \quad (54)$$

where $Z$ denotes the *total* number of the modes of the complete string in the frequency interval specified by the last formula in Eq. (52c). In obtaining Eq. (54) we have taken into account the elementary relation $I_{n,n} = l/2$. It is clear that in the present case – under the conditions listed in Eqs. (53a-b) – the ensemble average, Eq. (54), of the the energy $e(t)$ just obtained coincides with the time average, that is

$$\eta \equiv \{e(t)\} = e_0 , \quad (55)$$

where the curly bracket { } denotes time averaging. However, this is not true for the fluctuations, as Ehrenfest has shown. After a lengthy but straightforward calculation he obtained the following results using pure ensemble averaging

$$\frac{\overline{(e-e_0)^2}}{e_0^2} = \left(\frac{1}{z} - \frac{2}{Z}\right) + \frac{\overline{B^4}}{\overline{B^2}^2} \cdot \frac{1}{Z}, \text{ where } z \equiv \frac{l}{L}Z. \quad (56a)$$

Notice that in the above equation we have introduced the number of modes in the "sub-volume" denoted by $z$, which is only a fraction of the total number of modes $Z$ in the spectral range ($\omega'$,$\omega'+d\omega'$). Now, if we consider in Eq. (56) the factor coming from the statistics as the ratio of the expectation value of the energy squared to the squared of the expectation value of the energy,

$$\frac{\overline{B^4}}{\overline{B^2}^2} = \frac{\overline{\varepsilon^2}}{\overline{\varepsilon}^2}, \quad (56b)$$

then, by using the Planck-Bose distribution, Eq. (4), we obtain

$$\frac{\overline{(e-e_0)^2}}{e_0^2} = \frac{1}{\overline{n} \cdot Z} + \frac{1}{z}. \quad (57)$$

From the definition of $z$ given in Eq. (56a) it is clear that if the length of the string goes to infinity, i.e. $l / L \longrightarrow 0$, then in Eq. (57) the second term – which describes the wave-like fluctuations – dominates over the quantum term. On the other hand in Einstein's fluctuation formula, Eq. (48a) – where the $m_v$ corresponds to a fixed $z$ value – neither of the terms are sensitive to the size of the *Hohlraum*.

The relative energy fluctuations have a completely different form when we calculate them by first time-averaging and then taking the ensemble averages. Ehrenfest obtained

$$\frac{\overline{\{(e-\eta)^2\}}}{e_0^2} = \frac{1}{z} - \frac{1}{Z}, \quad (58)$$

where $\eta$ is given in Eq. (55). It is remarkable that this quantity does not depend on the statistics of the ensemble. Equation (58) can also be obtained formally from Eq. (56a) by using the factorization $<B^4> = <B^2>^2$ (the bracket $< >$ replaces in the text the upper dash in the numbered formulas). It is seen from Eqs. (48a-b) that if we incease the size of the sub-volume $\upsilon$ n-times – by keeping the total mode number $M_V$ and the energy of the *Hohlraum* fixed – the variance of the energy in $\upsilon$ increases by a factor of n. This is not true in neither of the cases considered by Ehrenfest, as one can conclude from Eqs. (57) and (58). This means that, due to wave interference, the fluctuations in the sub-volumes are not independent. One should keep in mind that the sub-volume $\upsilon$ may consist of spatially separated parts! The question naturally emerges: which of these two hypotheses fit better to the reality? The additivity of the entropy, or the present approach, which is based on interference of "quantized waves"? Ehrenfest concludes that the latter one should not be kept agains the former one.

According to Adolf Smekal [44], if one uses a statistics for the energy distribution of the spectral components of the normal modes of the *Hohlraum*, then one implicitly assumes the presence of "Planck's Kohlenstäubchen" (a small carbon particle) in the sub-volume, which not only *transforms* the original radiation into a true black-body radiation, but also continuously *rearranges* it in case of any changes of the boundaries. An ideal *Hohlraum* with perfectly reflecting walls is a completely deterministic system which is not able to rearrange itself according to classical electrodynamics, except for the case when we assume, *in addition*, that the eigenmodes themselves are autonomous objects – like the material particles – which function like the atoms, and are able to give each other energy spontaneously. So, if we refuse the latter additional assumption, then we have to accept that the radiation processes cannot take place without the assistance of material agents which are capable of absorbing, emitting and rearranging the energies of the different spectral components of the radiation.
If we write Ehrenfest's formula, Eq. (57), in the form

$$\overline{(e-\bar{e})^2} = h\nu \cdot \frac{\upsilon}{V}\bar{e} + \frac{c^3}{8\pi\nu^2 d\nu} \cdot \frac{\overline{e}^2}{\upsilon}, \tag{59}$$

we nicely see that if the boundaries of the Hohlraum are going to infinity ($V \longrightarrow 0$), then the quantum term goes to zero, and only the wave-like fluctuations survive. On the other hand, if we have a look at Einstein's fluctuation formula, Eq. (10), in this notation

$$\overline{(e-\bar{e})^2} = h\nu \cdot \bar{e} + \frac{c^3}{8\pi\nu^2 d\nu} \cdot \frac{\overline{e}^2}{\upsilon}, \tag{60}$$

then we see that in the limit $V \longrightarrow 0$ each term survives. Smekal argued that the difference of the two results, Eqs. (59) and (60), is quite understandable, because the two formulas refer to two different boundary conditions. *Namely, in the case considered by Ehrenfest the sub-volume $\upsilon$ is completely free from any material agents, on the other hand Einstein – since he used Planck's low for the spectral density – in fact, implicitely assumed the presence of a carbon particle in $\upsilon$. In the first case, if one lets the walls of the cavity going to infinity, then only the wave type fluctuations – stemming from beats due to interference – survive.* In this context one should keep in mind that the wave-like fluctuation (the second term on the rhs of Eqs. (59) or (60)) has always such a form for *any* (not only black-body) radiations, as was originally shown by Lorentz [45]. In the second case the quantum term survives the limit, since the presence of the carbon particles secures the continuous rearrangement of the energies of the spectral components by spontaneous and induced radiative transitions. In general, the exchange of the amount of energy $h\nu$ between two particular mode in the spectral range ($\nu$, $\nu+d\nu$) is possible only through the mediation of a material system capable of absorbing and emitting radiation at this frequency. Having absorbed the energy from one mode, the material system goes from state *1* to state *2*, and can emit the energy to the same or to another mode after a shorter or longer time. The probability of absorption is proportional with the product of the spectral density $u(\nu,T)$, Eq. (3), and the Einstein $B_{21}$ coefficient. The emission can take place, on one hand, spontaneously, characterized by the $A_{12}$ coefficient, and, on the other hand, by stimulated emission whose probability is proportional to the product of $u(\nu,T)$ and the second Einstein coefficient $B_{12}$ [46]. Smekal argues that the joint probability of this compound process is the product of the probabilities of absorption and of emission if we neglect the lifetime of the upper state *2* of the material system. The probability of such a double process is proportional to the expression

$$B_{21}u(\nu,T) \cdot [A_{12} + B_{12}u(\nu,T)] = B_{21}A_{12}u(\nu,T) + B_{21}B_{12}[u(\nu,T)]^2$$
$$= B^2 \frac{8\pi\nu^2}{c^3}\left\{h\nu \cdot u(\nu,T) + \frac{c^3}{8\pi\nu^2}[u(\nu,T)]^2\right\}. \tag{61}$$

In deriving Eq. (61) we have taken into account the following relations between the Einstein coefficients [46]

$$A_{12} = \frac{8\pi h\nu^3}{c^3}B_{12}, \text{ and } B_{21} = B_{12} \equiv B. \tag{62}$$

We note that, for simplicity, we have assumed the statistical weights (*a priori probabilities*) of the states *1* and *2* to be the same. By multiplying the expression in the curly bracket in Eq. (61) with the product of a sub-volume and the width of the spectral range, $\upsilon d\nu$, we exactly recover Einstein's fluctuation formula given by Eq. (60). According to Smekal, now the physical meaning of the two terms in the fluctuation formula is at hand. *The first term describing the particle-like fluctuations comes from the spontaneous processes of the* *radiation-matter interaction. The wave-like fluctuations (the so called interference term) are stemming from the induced processes* (absorption and "negative absorption" or, in the usual terminology; induced emission) of the material system taking part in the interaction. In an empty *Hohlraum* the stationary energy content of the radiation in a particular direction clearly means that this component is generated by pairwise counterpropagating waves. When we consider standing waves in a *Hohlraum* with reflecting walls facing opposite to each other – as was done first by Lorentz – then this pair of waves are coupled to each other in such a way, at least on average, that the same amount of radiation is being absorbed and reemitted in that particular direction. Thus each of the interacting agents (the material system and the radiation mode) change their energies by $h\nu$ periodically in an opposite manner, so that the energy of the mode does not change on average. Smekal characterized this situation with the word "Lichtquanten-Paternosterwerk". This means that the material of Lorentz's mirrors behave quasi-classically: the material constituents of real mirrors, of course, undergo spontaneous transitions, too, so the particle-like fluctuations are also present in the cavity. If the mode of the radiation stays on average on the energy ladder characterized by a very high excitation $n$, then the contribution from the spontaneous process can be neglected, and then in the fluctuation the interference term dominates. It is easy to realize that Smekal's "Lichtquanten-Paternosterwerk" is nothing else but a preliminary picture for the *Rabi oscillations*, whose detailed analysis can be found in any textbook on quantum electronics. We also note that – according to Smekal – the "communication" (energy exchange) of two modes with *different propagation directions and polarization* is possible only through the spontaneous transitions of the material system.

### 5.2. *Fluctuation from field quantization*

In their famous "Drei Männer Arbeit" (three men's work) Max Born, Werner Heisenberg and Pascual Jordan in 1925 – besides giving solutions to several fundamental problems, which constitute today basic parts of any standard texts on quantum mechanics – published a first preliminary version of field quantization [47], or in other words, the quantization of a continuous dynamical system. For simplicity, like Ehrenfest, they studied the small transverse oscillations of a string characterized by Eqs. (49a-b) given in the former subsection, and they quantized the amplitudes of the normal modes. In modern notation, they used the following expression for the quantized field $u(x,t)$

$$u(x,t) = \sum_{n=1}^{\infty}(\hbar/\omega_n)^{1/2} q_n(t) f_n(x), \text{ where } f_n(x) = (2/\sigma L)^{1/2}\sin k_n x, \tag{63a}$$

$$q_n(t) = (a_n e^{-i\omega_n t} + a_n^+ e^{+i\omega_n t})/2^{1/2}, \quad \omega_n = ck_n = n\cdot(c\pi/L), \tag{63b}$$

where we have introduced the creation and annihilation operators (matrices) $a_n$ and $a_n^+$ of the excitations of normal modes of the string. For later convenience we introduce the dimensionless conjugate momenta

$$p_n(t) = (a_n e^{-i\omega_n t} - a_n^+ e^{+i\omega_n t})/2^{1/2} i, \tag{63c}$$

and summarize the Heisenberg commutation rules in dimensionless form

$$[a_n, a_m^+] = \delta_{nm}, \quad [q_n(t), p_m(t)] = i\delta_{nm}. \tag{63d}$$

The total energy of the string is calculated analogously to the classical case, Eq. (50c)

$$H = (1/2)\int_0^L dx[\sigma(\partial u/\partial t)^2 + \tau(\partial u/\partial x)^2] = (1/2)\sum_{n=1}^{\infty}\hbar\omega_n(a_n^+ a_n + a_n a_n^+)$$
$$= \sum_{n=1}^{\infty}\hbar\omega_n\left(a_n^+ a_n + \frac{1}{2}\right) \tag{64}$$

We see that there is an infinite contribution coming from the zero-point energy $\hbar\omega_n/2$ of each mode. We mention here that the zero-point energy $h\nu/2$ of the material oscillators was first found by Planck [21-24] already in 1911. With the following replacements of the matrices by c-numbers

$$a_n \to (\omega_n/2\hbar)^{1/2} C_n e^{-i\phi_n} \;,\quad a_n^+ \to (\omega_n/2\hbar)^{1/2} C_n e^{+i\phi_n}, \tag{65a}$$

we recover the classical expression Eq. (50c) for the total energy of the string. The real classical dynamical variables take then the form,

$$(\hbar/\omega_n)^{1/2} q_n(t) \to q_n^{cl}(t) = C_n \cos(\omega_n t + \phi_n), \tag{65b}$$

$$(\hbar\omega_n)^{1/2} p_n(t) \to p_n^{cl}(t) = -\omega_n C_n \sin(\omega_n t + \phi_n) = dq_n^{cl}(t)/dt. \tag{65c}$$

As is well known, the matrices $a_n^+ a_n$ in Eq. (64) are diagonal possessing non-negative integer eigenvalues $N_n = 0,1,2,\ldots$ . Now let us calculate – similarly to Eq. (51) – the energy matrix of a segment $(0, l)$ of the string. We obtain

$$\begin{aligned}E &= \frac{l}{L}H + \sum_{n\neq m}\hbar\sqrt{\omega_n\omega_m}\,[q_n q_m K'_{nm} + p_n p_m K_{nm}]/2L \\ &+ \sum_n \hbar\omega_n(q_n^2 - p_n^2)\sin(2k_n l)/2k_n\end{aligned}, \tag{66a}$$

$$K'_{nm} = \frac{\sin(k_n - k_m)l}{k_n - k_m} + \frac{\sin(k_n + k_m)l}{k_n + k_m}, \tag{66b}$$

$$K_{nm} = \frac{\sin(k_n - k_m)l}{k_n - k_m} - \frac{\sin(k_n + k_m)l}{k_n + k_m}. \tag{66c}$$

Since we are interested in the fluctuations in a narrow spectral range whose wavelengths $\lambda_n = 2L/n$ are much smaller than both $L$ and $l$, the third term in Eq. (66a) can be supressed due to fast oscillations, so the matrix of the deviation of the energy from the diagonal part $(l/L)H$ (the first term on the rhs of Eq. (66a)) is given as

$$\Delta = \sum_{n\neq m}\hbar\sqrt{\omega_n\omega_m}\,[q_n q_m K'_{nm} + p_n p_m K_{nm}]/2L, \tag{67}$$

hence

$$\begin{aligned}\Delta^2 &= \Delta_1^2 + \Delta_2^2 + \Delta_1\Delta_2 + \Delta_2\Delta_1 = \\ &(1/4L^2)\sum_{n\neq m}\sum_{r\neq s}\hbar^2\sqrt{\omega_n\omega_m\omega_r\omega_s}\,\{[q_n q_m q_r q_s K'_{nm}K'_{rs} + p_n p_m p_r p_s K_{nm}K_{rs}] \\ &+ [q_n q_m p_r p_s K'_{nm}K_{rs} + p_n p_m q_r q_s K_{nm}K'_{rs}]\}\end{aligned} \tag{68}$$

According to Born, Heisenberg and Jordan, under the phase average of a matrix we will mean that diagonal matrix whose diagonals coincide with that of the original matrix. This "phase-averaging" is certainly justified in the case of thermal radiation.

For a comparison, first we study the classical case, and the variables $q$ and $p$ are considered to be c-numbers. By assuming uniform independent distributions for the phases in Eq. (65b-c), it can be shown from Eq. (68) that

$$\overline{\Delta_1^{cl}\Delta_2^{cl} + \Delta_2^{cl}\Delta_1^{cl}} = 0, \tag{69a}$$

$$\overline{(\Delta_1^{cl})^2 + (\Delta_2^{cl})^2} = (1/2L^2)\sum_{n,m=1}^{\infty}\{\omega_n^2\omega_m^2\,\overline{q_n^2}\,\overline{q_m^2}\,K'^2_{nm} + \overline{p_n^2}\,\overline{p_m^2}\,K^2_{nm}\}. \tag{69b}$$

Now we suppose that the length $L$ of the string is so large that the eigenfrequencies are spaced such densely that the sums in Eq. (69b) can be well approximated by integrals,

$$\overline{(\Delta_1^{cl})^2 + (\Delta_2^{cl})^2}$$
$$= \frac{1}{2\pi^2} \int_0^\infty d\omega \int_0^\infty d\omega' \left\{ \omega^2 \omega'^2 \overline{q_\omega^2 q_{\omega'}^2} + \overline{p_\omega^2 p_{\omega'}^2} \right\} \frac{\sin^2[(\omega-\omega')l/c]}{(\omega-\omega')^2}, \quad (69c)$$

where we have taken into account that in both $K_{nm}$ and $K'_{nm}$ in Eqs. (66b-c) the contribution of second, "non-resonant" term can be neglected. Because for very large $l$ the kernel in Eq. (69c) approximates the delta function,

$$\frac{c}{\pi \cdot l} \frac{\sin^2[(\omega-\omega')l/c]}{(\omega-\omega')^2} \to \delta(\omega-\omega'), \quad (69d)$$

finally we obtain

$$\overline{(\Delta_1^{cl})^2 + (\Delta_2^{cl})^2} = \frac{l}{2\pi \cdot c} \int_0^\infty d\omega \left\{ \omega^2 \omega^2 \overline{q_\omega^2 q_\omega^2} + \overline{p_\omega^2 p_\omega^2} \right\}$$
$$= \frac{l}{4\pi \cdot c} \int_0^\infty d\omega (\omega^2 \overline{C_\omega^2})^2 \quad . \quad (70)$$

As can be seen from Eq.(66a), the average energy $E$ of the segment $(0, l)$ of the string equals $(l/L)$ times the total energy, Eq. (50c). Going over from the summation to integration in Eq. (50c) we have

$$\overline{E} = \frac{l}{2\pi \cdot c} \int_0^\infty d\omega (\omega^2 \overline{C_\omega^2}). \quad (71)$$

By comparing Eqs. (70) and (71) we can easily express the fluctuation in a narrow spectral range $(\omega, \omega+d\omega)$,

$$\overline{(\Delta_\omega^{cl})^2} = \frac{c}{2d\nu} \cdot \frac{\overline{E}_\nu^2}{l}, \quad (72)$$

where we have introduced the average energy in that spectral range, and we have taken into account that $\omega = 2\pi\nu$. Eq. (72) is in complete analogy with the second term in Eq. (60) accounting for the "interference fluctuation". We note that the spectral mode density for the one-dimensional case under discussion is equal to $2/c$, hence the quantity $(2/c)ld\nu$ on the right hand side of Eq. (72) is just the number of modes in the segment $(0,l)$ in the spectral range $(\nu, \nu+d\nu)$.

Now we come back to the analysis of the original quantum expression given by Eq. (68). In complete analogy with Eq. (70) we obtain

$$\overline{\Delta_1^2 + \Delta_2^2} = \frac{l}{2\pi \cdot c} \int_0^\infty d\omega \hbar^2 \omega^2 \left\{ \overline{q_\omega^2 q_\omega^2} + \overline{p_\omega^2 p_\omega^2} \right\}$$
$$= \frac{l}{\pi \cdot c} \int_0^\infty d\omega \left\{ (\hbar\omega a_\omega^+ a_\omega)^2 + \hbar\omega(\hbar\omega a_\omega^+ a_\omega) + (\hbar\omega)^2/4 \right\}, \quad (73)$$

but now $q_\omega, p_\omega$ and $a_\omega, a_\omega^+$ denote the dimensionless quantum variables defined in Eqs. (63b-c), and the upper dash means quantum averages, i. e. diagonalization, as was already mentioned after Eq. (68). Similarly, for the last two sums in Eq. (68) we obtain

$$\overline{\Delta_1\Delta_2 + \Delta_2\Delta_2} = \frac{l}{2\pi \cdot c} \int_0^\infty d\omega \hbar^2 \omega^2 \left\{ \overline{(q_\omega p_\omega)^2} + \overline{(p_\omega q_\omega)^2} \right\}$$
$$= -\frac{l}{\pi \cdot c} \int_0^\infty d\omega \left\{ (\hbar\omega)^2/4 \right\} \quad . \quad (74)$$

We see that when we add Eqs. (73) and (74) then the last term on the rhs of Eq. (73) – coming from the zero-point energy – cancels, on the other hand, the second term survives. This latter term represents the particle-like fluctuations. If we calculate from the first term of Eq. (66a) the average energy matrix of the segment (0, *l*) of the string we have

$$E = \frac{l}{\pi \cdot c} \int_0^\infty d\omega (\hbar \omega a_\omega^+ a_\omega). \tag{75}$$

After this point Born, Heisenberg and Jordan implicitely uses a semiclassical assumption in order to obtain the fluctuation formula. They make the correspondence

$$\frac{l}{\pi \cdot c} d\omega \hbar \omega a_\omega^+ a_\omega \to \overline{E}_\nu, \tag{76}$$

where, according to Eq. (76), they have introduced the average energy of the segment of the string in a spectral range (*ν, ν+dν*). Then from the first two terms on the rhs of Eq. (73) they obtain

$$\overline{\Delta^2} = h\nu \overline{E}_\nu + \frac{c}{2 d\nu} \cdot \frac{\overline{E}_\nu^2}{l}, \tag{77}$$

in complete analogy with Einstein's fluctuation formula Eq. (60). As we see from Eq. (73), *the origin of the particle-like fluctuation is the presence of the zero-point energy term due to the non-commutativity of the quantized amplitudes. Hence, according to Born, Heisenberg and Jordan, the appearance of the particle-like fluctuation is a kinematic effect, inherently contained in the quantized nature of the field.* The results embodied in Eqs. (73) and (74) can symbolically be sketched for two nearly spaced modes as

$$\overline{\Delta^2} \sim (h\nu)^2 \overline{(a_\nu^+ a_\nu + 1/2)(a_{\nu'}^+ a_{\nu'} + 1/2)} - (h\nu)^2/4 \to h\nu \cdot \overline{E}_\nu + \overline{E}_\nu^2.$$

It is remarkable that Born, Heisenberg and Jordan considered the derivation of Einsein's fluctuation formula as one of the crutial test of quantum mechanics, and they viewed the result expressed by Eq. (77) as an important support for their new theory.

## 6. Derivation of Einstein's Fluctuation Formula from Pure Particle-Like Fluctuations and from Fermion Type Fluctuations

As we have already mentioned in sub-section 3.3, the sum of the two terms in Eq. (13), describing the complete fluctuation of a particular mode, can be expanded into an infinite sum, in which each term represents purely particle-like fluctuations, as is shown in Eq. (17). The root of this property is that a component of the heat radiation in a mode can be considered – at least from the point of view of its energy and statistics – as a mixture of infinitely many noninteracting ideal gases consisting of "*photo-molecules*" or "*photo-multiplets*" of energies $mh\nu$ with $m$ = 1, 2, 3, ... which follow the Boltzmann statistics. The thermodynamical independence of these ideal gases was shown by Wolfke [48] in 1921, and the corresponding fluctuation formula was derived by Bothe [49] in 1923. However neither of these works presents a systematic discussion of the *complete statistics* of the photo-molecules. In the followings we shall give an analysis of the Bose distribution on the basis of classical probability theory [52]. On one hand, we shall derive the complete statistics of the photo-molecules, proposed by de Broglie, Wolfke and Bothe. On the other hand we shall derive a new division of the Bose distribution into an infinite sum of Fermi distributions of "*binary photo-molecules*" of energies $2^s h\nu$, with $s$=0, 1, 2, ….

### 6.1. *The infinite divisibility of the Bose distribution: Poisson photo-molecules*
Let us consider the mode energy $\xi$ as a classical random variable of the discrete distribution $f_\xi(n)$ given by Eq. (4)

$$f_\xi(n) \equiv p_n \equiv P(\xi = n) = (1-b)b^n = \frac{1}{1+\bar{n}}\left(\frac{\bar{n}}{1+\bar{n}}\right)^n, \text{ where } b = e^{-h\nu/kT}. \tag{78}$$

Now we show that the Bose distribution is infinitely divisible. A random variable $\eta$ is said to be infinitely divisible if for any natural number $n$, it can be written down as a sum of completely independent random variables having the same distributions: $\eta = \eta_1+\eta_2+...+\eta_n$ [50]. The infinite divisibility of a distribution can be conveniently studied with the help of its characteristic function [51], because in this case the characteristic function of the sum variable equals to the product of the characteristic functions of the summands. The Fourier transform of the distribution Eq. (78) reads

$$\varphi_\xi(t) = \langle e^{i\xi \cdot t}\rangle = (1-b)\sum_{n=0}^{\infty} b^n e^{in\cdot t} = \frac{1-b}{1-be^{it}}. \tag{79}$$

The logarithm of this characteristic function can be expanded into the power series

$$\log[\varphi_\xi(t)] = \log\left(\frac{1}{1-be^{it}}\right) - \log\left(\frac{1}{1-b}\right) = \sum_{m=1}^{\infty} \frac{b^m}{m}\left(e^{imt}-1\right), \tag{80}$$

$$b = e^{-h\nu/kT} < 1,$$

where each term is a logarithm of the characteristic function of Poisson distributions with parameters $b^m/m$, where $m = 1,2, \ldots$ are the multiplet indeces (the number of energy quanta $h\nu$ in the photo-molecules). This means that the characteristic function can be properly factorized, and the random variable $\xi$ itself is represented by an infinite sum,

$$\varphi_\xi(t) = \varphi_{x_1}(t)\varphi_{x_2}(t)\cdots\varphi_{x_m}(t)\cdots, \tag{81}$$

$$\xi = x_1 + x_2 + ... + x_m + ..., \tag{82}$$

The independent random variables $x_m$ have the Poisson distributions

$$p_l(m) = P(x_m = l \cdot m) = \frac{\lambda_m^l}{l!}e^{-\lambda_m}, \quad \lambda_m \equiv \frac{b^m}{m} = \frac{e^{-mh\nu/kT}}{m}. \tag{83}$$

The average energy of the $m$-th multiplet is given by

$$E_m \equiv \overline{E}_m = h\nu\overline{x_m} = h\nu \cdot m \cdot \lambda_m = h\nu \cdot b^m = h\nu \cdot e^{-mh\nu/kT}. \tag{84}$$

Because the variance of the Poisson distribution is equal to its parameter, the fluctuation of the energies of the multiplets read

$$\Delta E_m^2 = (h\nu)^2 \Delta x_m^2 = (h\nu)^2 m^2 \lambda_m = mh\nu \overline{E}_m. \tag{85}$$

According to Eq. (85) the energies of the photo-molecules have only particle-like fluctuations. As can be easily seen, the sum of the contributions, Eq. (85), gives back the complete fluctuation, Eq. (13), as has already been written down in Eq. (17)

$$\Delta E^2 = (h\nu)^2(\bar{n}+\bar{n}^2) = (h\nu)^2 \Delta\xi^2 = (h\nu)^2 \sum_{m=1}^{\infty} \Delta x_m^2 = \sum_{m=1}^{\infty} \Delta E_m^2. \tag{86}$$

The entropy $S_m$ of the $m$-th photo-multiplet gas is obtained by integrating the thermodynamic relation

$$dS_m/dE_m = 1/T = -(k/mh\nu)\log(E_m/h\nu), \tag{87}$$

$$\begin{aligned}S_m &= k[(E_m/mh\nu) - (E_m/mh\nu)\log(E_m/h\nu)] \\ &= (k/m)[\overline{x_m} - \overline{x_m}\log\overline{x_m}]\end{aligned} \tag{88}$$

According to Eqs. (1) and (2), the entropy of a mode of the black-body radiation can be recovered as a sum of the entropies, Eq. (88), of the photo-multiplet gas components [48]

$$S = k[(1+\bar{n})\log(1+\bar{n}) - \bar{n}\log\bar{n}] = \sum_{m=1}^{\infty} S_m . \tag{89}$$

This means that the photo-multiplet gas components are thermodynamically independent.

Let us now introduce the total energy $E(m)$ of the $m$-th multiplet gas in the frequency range $(\nu, \nu+d\nu)$ in a volume $V$,

$$E(m) = VZ_\nu d\nu E_m , \quad \text{where} \quad Z_\nu = \frac{8\pi\nu^2}{c^3} . \tag{90}$$

The corresponding entropy is obtained by using Eq. (88),

$$S(m) = k[E(m)/mh\nu]\{1 - \log[E(m)/h\nu V Z_\nu d\nu]\} . \tag{91}$$

Following Einstein's argumentation [3] – as was presented in sub-section 3.1 – we calculate from Eq. (91) the entropy difference of two states of the $m$-th multiplet gas having the same energy but occupying the different volumes $V$ and $V_0 > V$:

$$S(m) - S_0(m) = k\log\left[\left(\frac{V}{V_0}\right)^{\frac{E(m)}{mh\nu}}\right] = k\log\left[\left(\frac{V}{V_0}\right)^l\right], \tag{92}$$

where the last equation in Eq. (92) contains the geometrical probability $w^l = (V/V_0)^l$ for $l$ independent particles occupying the part $V$ of a larger volume $V_0$. According to Eq. (92) we may say – in complete analogy with Einstein's original statement – that *a monochromatic multiplet component (in the frequency range ($\nu, \nu+d\nu$)) of a black-body radiation – from the point of view of thermodynamics – behaves like a classical gas, as if it consisted of independent energy quanta $mh\nu$* [48]. This is a generalization of Einstein's introduction of light quanta of energy $h\nu$ (which is a special case with $m = 1$). His analysis was restricted to the Wien limit, $h\nu/kT \gg 1$, which is an approximation to Planck's exact formula for small radiation densities. The above generalization is valid for any values $0 < h\nu/kT < \infty$. Einstein's argumentation cannot be applied by a direct use of Planck's entropy expression Eq. (1). However, by introducing the photo-multiplets, Einstein's reasoning can be generalized, as is shown by Eq. (92). For small radiation densities the first order "muliplet" component dominates (these are Einstein's original light quanta), but for large densities the higher order multiplets become more and more important. We note that this is one possible way to recover the Rayleigh-Jeans formula (in the limit $h\nu/kT \ll 1$) from the photon concept (i. e. from the particle description). We note that Bothe has shown [49] that the Planck formula can be deduced from the separate conservation of the number of photo-multiplets when they interact with a two level material system. The higher order multiplets take part only in the induced processes; when we leave out their contributions we arrive at the Wien formula. According to the above analysis, *the black-body radiation in any narrow spectral range can be considered as a mixture of infinitely many statistically and thermodynamically independent classical photon gases consisting of "photo-molecules" or "photo-multiplets" of energies $h\nu, 2h\nu, 3h\nu, ..., mh\nu, ....* We note that each Poisson component can be divided further to Poisson variables again – that is, the Poisson distribution is not an irreducible distribution – but, as far as we know, no physical interpretation can be attached to these variables.

### 6.2. *The infinite divisibility of the Bose distribution: Binary photo-molecules*

In the present sub-section we prove that the Bose variable whose distribution is given by Eq. (78), can be decomposed into an infinite sum of binary random variables, which correspond to "binary photo-molecules" (we may also call them "fermion photo-multiplets", or simply

"binary photons") containing $2^s = 1, 2, 4, 8, \ldots$ single photon energies, where $s = 0, 1, 2, \ldots$ [52].

By using the algebraic identity

$$(1-z)(1+z)(1+z^2)\cdot\ldots\cdot(1+z^{2^s}) = 1 - z^{2^{s+1}} \tag{93a}$$

we have the following absolutely convergent infinite product representation of $1/(1-z)$

$$\frac{1}{1-z} = \prod_{s=0}^{\infty}(1+z^{2^s}) \qquad (|z|<1). \tag{93b}$$

With the help of Eq. (93b) the characteristic function of the Bose distribution of the random variable $\xi$, Eq. (79), can be expanded into the infinite product of characteristic functions having similar functional forms [51]:

$$\varphi_\xi(t) = \frac{1-b}{1-be^{it}} = \prod_{s=0}^{\infty}\frac{1+b^{2^s}e^{2^s it}}{1+b^{2^s}} = \frac{1+be^{it}}{1+b}\cdot\frac{1+b^2 e^{2it}}{1+b^2}\cdot\frac{1+b^4 e^{4it}}{1+b^4}\cdot\ldots$$

$$= \prod_{s=0}^{\infty}\varphi_{u_s}(t), \quad (b = e^{-h\nu/kT}) \tag{94}$$

Hence, the random variable $\xi$ can be decomposed into an infinite sum of independent variables $\{u_s, s = 0, 1, 2, \ldots\}$,

$$\xi = u_0 + u_1 + u_2 + \ldots + u_s + \ldots, \tag{95}$$

which have the binary distributions

$$p_0(s) = P(u_s = 0) = \frac{1}{1+b^{2^s}}, \qquad p_1(s) = P(u_s = 2^s) = \frac{b^{2^s}}{1+b^{2^s}} \tag{96}$$

($s = 0, 1, 2, \ldots$).

The multiplets just obtained have occupation numbers 0 or 1, hence they formally follow the exclusion principle, which means that they behave like fermions. One can easily check that the characteristic function of these variables are really the factors of the infinite product Eq. (94)

$$\varphi_{u_s}(t) = \langle e^{iu_s t}\rangle = \frac{1+b^n e^{in\cdot t}}{1+b^n} \qquad (n \equiv 2^s). \tag{97}$$

The decomposition Eq. (95) is an irreducible decomposition, i. e. the variables $u_s$ cannot be divided further. The expectation value of the energy of the $s$-th fermion multiplet becomes

$$E_s \equiv \overline{E}_s = h\nu\overline{u}_s = 2^s h\nu\frac{1}{\exp(2^s h\nu/kT)+1} = 2^s h\nu\cdot\overline{n}_s, \tag{98a}$$

where

$$\overline{n}_s = \frac{1}{\exp(2^s h\nu/kT)+1} \tag{98b}$$

denotes the mean occupation number. The expression in Eq. (98a) is a Fermi distribution with zero chemical potential. The fluctuation of the $s$-th multiplet reads

$$\Delta E_s^2 = (h\nu)^2 \Delta u_s^2 = (2^s h\nu)^2(\overline{n_s^2} - \overline{n}_s^2) = 2^s h\nu E_s - E_s^2. \tag{99}$$

Eq. (99) shows that the energy fluctuation of the "binary photons" contain both particle-like and wave-like fluctuations, but the wave-term has a negative sign (which is a characteristics of fermion fluctuations) in contrast to the case of bosons. It can also be checked by direct calculation, that the sum of the contributions, Eq. (99), gives back the complete fluctuation, Eq. (13),

$$\Delta E^2 = (h\nu)^2 (\bar{n} + \bar{n}^2) = (h\nu)^2 \Delta \xi^2 = (h\nu)^2 \sum_{s=0}^{\infty} \Delta u_s^2 = \sum_{s=0}^{\infty} \Delta E_s^2 . \tag{100}$$

The entropies of the binary photon gas components can be obtained in several different ways. Here we use the usual Boltzmann definition:

$$S_s = -k\{p_0(s)\log[p_0(s)] + p_1(s)\log[p_1(s)]\}$$
$$= -k\{[1-(E_s/2^s h\nu)]\log[1-(E_s/2^s h\nu)] + (E_s/2^s h\nu)\log(E_s/2^s h\nu)\} .$$
$$= -k[(1-\bar{n}_s)\log(1-\bar{n}_s) + \bar{n}_s \log \bar{n}_s]$$
$$\tag{101}$$

From Eqs. (101) and (98b) it can be proved that the usual thermodynamic relation

$$dS_s / dE_s = 1/T \tag{102}$$

is satisfied for any $s$ values as an identity, so the binary photon gas components have the same temperature. The sum of the entropies of the binary photon gas components, Eq. (101), give back the Planck entropy of a mode of the black-body radiation, Eq. (1) (here Eq. (2) should also be taken into account):

$$S = k[(1+\bar{n})\log(1+\bar{n}) - \bar{n}\log\bar{n}] = \sum_{s=0}^{\infty} S_s . \tag{103}$$

Eq. (103) shows that the binary photon gas components are thermodynamically independent. We see from the above analysis, that the black-body radiation in any narrow spectral range can be considered as a mixture of infinitely many statistically and thermodynamically independent fermion "binary photon gases" consisting of "binary photo-molecules" or "fermion photo-multiplets" of energies hν, 2hν, 4hν,...,2$^s$hν,... . According to Eq. (100) and (99), Einstein's formula for the fluctuation of the energy of black-body radiation in a sub-volume of a Hohlraum can be expressed as a sum of fermion type fluctuations of the energies of the binay photon gas components. For small radiation densities (i. e. when hν/kT>>1) the zeroth order multiplet dominates, and we get back the Wien formula for the spectral density. In the opposite case – in the Rayleigh-Jeans limit – all the higher terms have to be taken into account in order to get back the correct asymptotic behaviour of the black-body spectrum. It is an open question whether the above decompositions have any physical significance at all. Nevertheless, the binary photons provide us with a natural basis for the dyadic expansion of an arbitrary incoherent excitation of a mode of the black-body radiation. For instance, for the 9-th excitation we have 9=1001 in the binary system, which means that the zeroth ($2^0$=1) and the third ($2^3$=8) binary photons have occupation number 1, and all the others have occupation number 0.

## 7. Summary

In the preceding sections we gave an overview on the early results related to Einstein's fluctuation formula and wave-particle duality. Our motivation has been to follow – on the basis of original publications – the development of the concept of light quanta, mostly in the context of black-body radiation, from its birth up to the work by Born, Heisenberg and Jordan. The latter contained the first example of field quantization. Having discussed the universal character of black-body radiation and Planck's discovery of the elementary quantum of action, in Section 2, we presented Einstein's arguments leading to the introduction of light quanta in Section 3. Then, following Einstein, we derived the fluctuation formula which contains both the "particle-term" and the "wave-term", and gave a first approach to the physical interpretation of these terms. The particle-term comes from the Poisson distribution of spatially localized light-quanta, and the wave term represents the random amplitude and phase fluctuations of the classical chaotic field. Since Einstein's fluctuation formula contains both terms, it can be considered as the first precise mathematical formulation of the wave-particle duality of light.

In Section 4 we presented the combinatorial analysis published by Natanson in 1911 which directly led to the Bose distribution of photon occupation numbers. We pointed out that from this analysis not only Bose's result – which was published 13 years later in 1924 – but also the Fermi distribution comes out quite naturally. In this context we also have to remark that the idea of quantization of energy dates back to the works of Boltzmann [53],[54] appeared in 1872 and 1877, respectively. Though Boltzmann considered the concept of discrete energy elements as a mere mathematical device, nevertheless in his second paper – on the basis of combinatorial analysis and maximization of the probability – he calculated the (Bose) distribution of the energy elements among the molecules of a gas. As Planck remarked on the Solvay Meeting in 1911 [23], the maximum number of combinations determined by Boltzmann gives the same entropy as the total number of them, because the relative difference of these two numbers are negligible. In fact Boltzmann's method was adapted by Natanson to derive the spectrum of black-body radiation, and this method was later rediscovered by Bose. In the second part of Section 4 we have briefly discussed a work by Planck on the fluctuations of classical fields. In this description the wave-like fluctuation can also be viewed as a result of an irregular breeding of the beat waves formed from the interfering Fourier components. We have seen there that, interestingly, a particle-like term can also be derived if we assume certain definite phase relations among the Fourier components of the field.

In Section 5 first we dealt with the results of Ehrenfest and Smekal who brought a new element into the discussion, namely the clear distinction of the following two cases: the study of the fluctuations in the absence of a material body in the sub-volume of the cavity, on one hand, and the the study of the fluctuation in the presence of a ponderable piece of matter (Planck's "Kohlenstäubchen", a small carbon particle), on the other hand. In the first case only the wave term survives in the large cavity limit. In the second case both terms survive, as in Einstein's original fluctuation formula. Ehrenfest and Smekal claimed that in Einstein's derivation the presence of a material body in the sub-volume had been implicitly taken into account. The black-body radiation and a material gas in thermal equilibrium differ fundamentally in that respect that the "molecular chaos" in the gas is caused by the continuous random collisions of the molecules. On the other hand, the modes of the radiation (or the photons, or von Laue's bundles of rays) do not interact directly (here we can safely leave out of consideration the "exotic" process of photon-photon scattering of high energy quantum electrodynamics), they cannot collide and proceed by themselves to an equilibrium with a maximum entropy. The energy transfer from one mode to another can take place only through the mediation of a piece of ponderable matter, which absorbs and emits radiation, and from time to time rearranges the energy distribution in such a way that a stable thermal equilibrium is reached. According to Smekal the particle-term comes from the spontaneous transitions of the material system (which makes the "communication" of modes, having different propagating direction and polarization, possible), and the appearance of the wave term is a result of induced processes (which do not change the propagation direction). In the second part of Section 5 we presented the classic work of Born, Heisenberg and Jordan on energy fluctuations, which contains the first example of the quantization of a continuum, where they represented the field amplitudes by matrices (operators) at each points of a string with fixed end-points. According to their analysis, the appearance of the particle-term in the fluctuation formula is a kinematic effect and it has an intimate connection with the zero-point energy due to the non-commutativity of the quantized amplitudes. The thermal character of the continuum was taken into accont by phase-averaging the matrices, a procedure on which they ment the supression of the off-diagonal elements in the fluctuation matrix.

In Section 6 we presented our recent results on the Bose distribution. On the basis of classical probability theory, we have shown that the Bose distribution is infinitely divisible in two ways. It can be decomposed into Poisson distributions of classical photo-molecules, and also into irreducible binary distributions corresponding to fermion type photo-molecules. According to these results, the black-body radiation in any narrow spectral range can be considered as a mixture of infinitely many statistically and thermodynamically independent classical photon gases consisting of "photo-molecules" or "photo-multiplets" of energies $h\nu, 2h\nu, 3h\nu,..., mh\nu,...$. Within the framework of this interpretation Einstein's original arguments on the photon concept (sub-section 3.1) can be generalized as has been shown first by Wolfke in 1921. This way the limitation to Wien's approximate formula is not needed. In the second part of Section 6 we have seen that the black-body radiation in any narrow spectral range can also be considered as a mixture of infinitely many statistically and thermodynamically independent fermion "binary photon gases" consisting of "binary photo-molecules" or "fermion photo-multiplets" of energies $h\nu, 2h\nu, 4h\nu,..., 2^s h\nu,...$. In this way the boson fluctuations are split into a sum of irreducible fermion fluctuations.

Concerning fluctuations in general, we have to mention that Einstein in 1924 – being the first who have applied Bose's statistics to material particles [55], [56] – derived a fluctuation formula for the number of particles of an ideal gas in a sub-volume of a container which communicates with the rest of the container through a narrow "energy window". The two-term expression so obtained is completely analogous to his original formula for the black-body radiation, thus the wave-particle duality has been shown to be a characteristics of the material particles, too. He explained the appearance of the interference term on the basis of de Broglie's theory on matter waves, which appeared just in 1924. The complete theory of an ideal gas based exlusively on de Broglie waves was worked out by E. Schrödinger [57] one year later. Finally, it is interesting to note that the general formula for the fluctuation of the energy (Eq. (12) of sub-section 3.2), $\Delta E^2 = kT^2 dE/dT$ – which is widely believed to be derivable only from statistical physics – can also be obtained on the basis of the Second Law of phenomenological thermodynamics, as was shown by Szilard [58] in 1925.

**Acknowledgements**

This work has been supported by the Hungarian National Scientific Research Foundation (OTKA, grant number T048324). We thank Prof. G. J. Székely of Bowling Green State University for drawing our attention to the mathematical literature on the irreducible decomposition of the discrete exponential distribution.

**References**

[1] A. Einstein, *Zum gegenwärtigen Stand des Strahlungsproblems, Phys. Zeitschr.* **10** (1909) 185-193.
[2] M. Planck, *Ueber das Gesetz der Energieverteilung im Normalspektrum, Ann. der Phys.* **4** (1901) 553-563. Appeared first : M. Planck, *Zur Theorie des Gesetzes der Energieverteilung im Normalspectrum, Verhandlungen der Deutsch. Phys. Ges*. **2** (1900) 237-245. (Sitzung vom 14. December 1900.)
[3] A. Einstein, *Über einen die Erzeugung und Verwandlung des Lichtes betreffenden heuristischen Gesichtspunkt, Ann. der Phys.* **17** (1905) 132-148.

[4] M. Planck, *Theorie der Wärmestrahlung,* Johann Ambrosius Barth-Verlag, Leipzig (1906/66 ), see also M. Planck, *The theory of heat radiation*, Dover Publications, Inc., New York (1959), this is a translation of the second edition of Planck's book by M. Masius from 1914.

[5] R. von Kövesligethy, *On the theory of continuous spectra*, Proc. Hung. Acad. Sci. Vol. XII. No. 11. (1885) 1-32. (in Hungarian). On page 30 a formula equivalent to the $\lambda_m T = const$ displacement law can be found. See also ; *Grundzüge einer theoretischen Spekralanalyse*, Druck und Verlag von H. W. Schmidt, Halle a/S. (1890). Formula (144) on page 169 expresses the displacement law.

[6] Sir Edmund Whittaker, *A History of the Theories of Aether and Electricity. II. The Modern Theories*, Thomas Nelson and Sons Ltd, London (1953)

[7] O. Lummer and E. Pringsheim, *Kritisches zur schwarzen Strahlung, Ann. der Phys.* **6** (1901) 192-210.

[8] H. Rubens und F. Kurlbaum, *Anwendung der Methode der Reststrahlen zur Prüfung des Strahlungsgesetzes, Ann. der Phys.* **2** (1901) 649-666.

[9] W. Wien, *Ueber die Energievertheilung im emissionsspektrum eines schwarzen Körpers, Ann. der Phys. und Chem.* **58** (1896) 662-669.

[10] M. Planck, *Ueber eine Verbesserung der Wien'schen Spectralgleichung, Verhandlungen der Deutsch. Phys. Ges*. **2** (1900) 202-204. (Sitzung vom 19. October 1900.)

[11] J. H. Jeans, *Zur Strahlungstheorie, Phys. Zeitschr.* **9** (1908) 853-855.

[12] M. Planck, *Ueber irreversible Strahlungsvorgänge, Ann. der Phys.* **1** (1900) 69-122.

[13] A. Einstein, *Zur Elektrodynamik bewegter Körper, Ann. der Phys.* **17** (1905) 891-921.

[14] G. N. Lewis, *The conservation of photons, Nature* **118** (1926) 874-875.

[15] P. Lenard, *Ueber die lichtelektrische Wirkung. Ann. der Phys.* **8** (1902) 149-198.

[16] M. von Laue, *Die Einsteinschen Energieschwankungen, Verhandlungen der Deutsch. Phys. Ges*. **17** (1915) 237-245. ( Sitzung vom 14. Mai 1915. )

[17] R. Kubo, H. Ichimura, T. Usui and N. Hashitsume, *Statistical Mechanics*, North-Holland, Amsterdam (1965)

[18] A. Rényi, *Wahrscheinlichkeitsrechnung mit einem Anhang über Informationstheorie*, VEB Deutsche Verlag der Wissenschaften, Berlin (1962)

[19] W. Feller, *An Introduction to Probability and Its Applications I., II.*, John Wiley & Sons Inc. New York (1970, 1971)

[20] S. I. Wawilow, *Die Mikrostruktur des Lichtes*, Akademie-Verlag, Berlin (1954)

[21] M. Planck, *Eine neue Strahlungshypothese, Verhandlungen der Deutsch. Phys. Ges*. **13** (1911) 138-148. ( Sitzung vom 3. Februar 1911. )

[22] M. Planck, *Zur Hypothese der Quantenemission, Sitzungsberichte der Preuß. Akad. Wiss*. (1911) 723-731.

[23] M. Planck, *Die Gesetze der Wärmestrahlung und die Hypothese der elmentaren Wirkungs-quanten*, Solvay-Kongress Brüssel 1911. Gauthier-Villars, Paris (1912); *Abhandlungen der Bunsenges*. **3** No. 7 (1913) 77-94.; W. Knapp, Halle a. S. (1914)

[24] M. Planck, *Über die Begründung des Gestezes der schwarzen Strahlung, Ann. der Phys.* (4) **37** (1912) 642-656.

[25] M. Planck, *Eine veränderte Formulierung der Quantenhypothese, Sitzungsberichte der Preuß. Akad. Wiss*. (1914) 918-923.

[26] M. Planck, *Bemerkung zur Quantenstatistik der Energieschwankungen, Sitzungsberichte der Preuß. Akad. Wiss*. (1923) 355-358.

[27] M. Planck, *Die Energieschwankungen bei der superposition periodischer Schwingungen, Sitzungsberichte der Preuß. Akad. Wiss*. (1923) 350-354.

[28] M. Planck, *Über die Natur der Wärmestrahlung*, Ann. der Phys. (4) **73** (1924) 272-288.

[29] A Einstein, *La Théorie du Rayonnement et les Quanta, Rapports et Discussions de la réunion tenue à Bruxelles*, du 30 octobre au 3 novembre 1911 sous auspices de M. E. Solvay, Paris (1912) p. 419.

[30] A. Einstein und L. Hopf, *Über einen Satz der Wahrscheinlichkeitsrechnung und seine Anwendung in der Strahlungstheorie*, Ann. der Phys. (4) **33** (1910) 1096-1104.

[31] A. Einstein und L. Hopf, *Statistische Untersuchung der Bewegung eines Resonators in einem Strahlungsfeld*, Ann. der Phys. (4) **40** (1913) 560.

[32] A. Einstein und O. Stern, *Einige Argumente für die Annahme einer molekularen Agitation beim absoluten Nullpunkt*, Ann. der Phys. (4) **33** (1910) 1105-1115.

[33] A. Einstein, *Über die Entwicklung unserer Anschauungen über das Wesen und Konstitution der Strahlung*, Phys. Zeitschr. **10** (1909) 817-826. Vorträge und Diskussionen von der 81. Naturforscherversammlung zu Salzburg, 81. Versammlung deutscher Naturforscher und Ärtzte zu Salzburg, vom 21. bis 25. September 1909.

[34] H. A. Lorentz, *Alte und neue Frage der Physik*, Phys. Zeitschr. **11** (1910) 1234-1257.

[35] L. Natanson, *Über die statistische Theorie der Strahlung*, Phys. Zeitschr. **11** (1911) 659-666. Translated from the original : *On the Statistical Theory of Radiation, Bulletin de l'Académie des Sciences de Cracovie* (A) (1911) 134-148.

[36] S. N. Bose, *Plancks Gesetz und Lichtquantenhypothese*, Zeitschr. für Phys. **26** (1924) 178-181.

[37] S. Jacobson, *Radiation fluctuations and thermal equilibrium*, Phys. Rev. **30** (1927) 936-943.

[38] S. Jacobson, *Note on the fluctuations in black body radiation*, Phys. Rev. **30** (1927) 944-947.

[39] P. Ehrenfest, *Energieschwankungen im Strahlungsfeld oder Kristallgitter bei superposition quantisierter Eigenschwingungen*, Zeitschr. für Phys. **34** (1925) 362-373.

[40] P. Debye, *Der Wahrscheinlichkeitsbegriff in der Theorie der Strahlung*, Ann. der Phys. (4) **33** (1910) 1427-1434.

[41] L. S. Ornstein und F. Zernike, *Energiewisselingen der zwarte starling en licht-atomen*, Akad. v. Wetensch. Amsterdam **28** (1919/20) 280.

[42] P. Ehrenfest, *Bemerkungen betreffs zweier Publikationen über Energieschwankungen*, Zeitschr. für Phys. **34** (1925) 316.

[43] A. L. Fetter and J. D. Walecka, *Theoretical mechanics of particles and continua*, McGraw-Hill Book Company, New York (1980)

[44] A. Smekal, *Zur Quantenstatistik der Hohlraumstrahlung und ihrer Wechselwirkungen mit der Materie*, Zeitschr. für Phys. **37** (1926) 319-341.

[45] H. A. Lorentz, *Les théories statistiques en thermodynamique*, Teubner, Leipzig (1916), IX.

[46] A. Einstein, *Zur Quantentheorie der Strahlung*, Phys. Zeitschr. **18** (1917) 121-128.

[47] M. Born, W. Heisenberg und P. Jordan, *Zur Quantenmechanik. II.*, Zeitschr. für Phys. **35** (1925) 557-615.

[48] M. Wolfke, *Einsteinsche lichtquanten und räumliche Struktur der Strahlung*, Phys. Zeitschr. **22** (1921) 375-379.

[49] W. Bothe, *Die räumliche Energieverteilung in der Hohlraumstrahlung*, Zeitschr. für Phys. **20** (1923) 145-152.

[50] B. V. Gnedenko and A. N. Kolmogorov, *Limit distributions for sums of independent random variables*, Addison-Wesley Publishing Co., Cambridge (Mass.) (1954)

[51] E. Lukács, *Characteristic functions*, Hafner Publishing Co., New York (1960)

[52] S. Varró, *in preparation*

[53]   L. Boltzmann, *Weitere Studien über das Wärmegleichgewicht unter Gasmolekülen*, *Sitzungsberichte der math.-naturwiss. Classe der kaiser. Akad. d. Wiss. Wien* (Sitzung am 10. October 1872.) LXVI. Bd. II. Abth. (1872) 275-370.

[54]   L. Boltzmann, *Über die Beziehung zwischen dem zweiten Hauptsatze der mechanischen Wärmetheorie und der Wahrscheinlichkeitsrechnung, respective den Sätzen über das Wärmegleichgewicht, Sitzungsberichte der math.-naturwiss. Classe der kaiser. Akad. d. Wiss. Wien* (Sitzung am 10. October 1872.) LXXVI. Bd. II. Abth. (1877) 373-435.

[55]   A. Einstein, *Quantentheorie des einatomigen idealen Gases, Sitzungsberichte der Preuß. Akad. Wiss*. (Gesamtsitzung vom 10. Juli 1924.) XXII. (1924) 261-267.

[56]   A. Einstein, *Quantentheorie des einatomigen idealen Gases. Zweite Abhandlung, Sitzungsberichte der Preuß. Akad. Wiss*. (Sitzung der phys.-math. Klasse vom 8. Januar 1925.) XXIII. (1925) 3-14.

[57]   E. Schrödinger, *Zur Einsteinsche Gastheorie*, *Phys. Zeitschr.* **27** (1921) 95-101.

[58]   L. Szilard, *Über die Ausdehnung der phänomenologischen Thermodynamik auf die Schwankungserscheinungen, Zeitschr. für Phys.* **32** (1925) 753-788. English translation: L. Szilard, *On the extension of phenomenological thermodynamics to fluctuation phenomena*, in The Collected Works of Leo Szilard. Scientific Papers, pp.70-102, MIT Press, London (1972)